\documentclass[
aps,onecolumn,preprintnumbers,superscriptaddress,nofootinbib]{revtex4}
\usepackage{graphicx}
\usepackage{amssymb}
\usepackage{color}
\usepackage{mathtools}
\usepackage{enumerate}

\newcommand{\be}{\begin{eqnarray}}
\newcommand{\ee}{\end{eqnarray}}

\newcommand{\beaa}{\begin{equation} \begin{array}{ll}}
\newcommand{\eeaa}{\end{equation} 	\end{array} }

\usepackage{graphicx}
\usepackage{mathrsfs}
\usepackage{float}
\usepackage{multirow}
\usepackage{booktabs}

\usepackage{float}

\begin{document}

\title{Geometric origin of the galaxies' dark side}

\author{Leonardo Modesto}
\email{lmodesto@sustech.edu.cn} 
\author{Tian Zhou}
\email{11930538@mail.sustech.edu.cn} 
\author{Qiang Li}
\email{qliphys@gmail.com} 
\affiliation{Department of Physics, Southern University of Science and Technology, Shenzhen 518055, China}


\begin{abstract}
We hereby show that Einstein's conformal gravity is able to explain simply on the geometric ground the galactic rotation curves without need to introduce any modification in both the gravitational as well as in the matter sector of the theory. Our result applies to any Weyl conformal invariant theory that admits the Schwarzschild's metric as an exact solution. However, we here mostly consider the Einstein's scalar-tensor theory, which is surely ghost-free. 
The geometry of each galaxy is described by a metric obtained making a singular rescaling of the Schwarzschild's spacetime. 
The new exact solution, which is asymptotically Anti-de Sitter, manifests an unattainable singularity at infinity that can not be reached in finite proper time, namely, the spacetime is geodetically complete.  
It deserves to be notice that we here think different from the usual. Indeed, instead of making the metric singularity-free, we make it apparently but harmlessly even more singular then the Schwarzschild's one. 
Finally, it is crucial to point that the Weyl's conformal symmetry is spontaneously broken to the new singular vacuum rather then the asymptotically flat Schwarzschild's one. The metric is unique according to: the null energy condition, the zero acceleration for photons in the Newtonian regime, and the homogeneity of the Universe at large scales. 
Once the matter is conformally coupled to gravity, the orbital velocity turns out to be asymptotically constant consistently with the observations and the Tully-Fisher relation. Hence, in the properly identified effective Newtonian theory, we consider the effect of all stars in a galaxy on a probe star to finally get the physical velocity profile that we compare with the observations in order to fit the only free parameter in the metric and the mass to luminosity ratio for each galaxy. 
Our fit is based on a sample of $175$ galaxies and shows that our velocity profile very well interpolates the galactic rotation-curves data for the most of spiral galaxies. The fitting results for the the mass to luminosity ratio turn out to be close to $1$ consistently with the absence of dark matter.

\end{abstract}

\maketitle

\tableofcontents

\section{Introduction}

Despite the enormous successes of the Einstein's theory of gravity, the latter appears to be about ``twenty five percent wrong''. So far the scientists proposed two possible solutions of the problem that are known under the name of ``dark matter'' or ``dark gravity", and both are extensions of the Einstein's field equations. The first proposal consists on modifying the right side of the Einstein's equations, while according to the second proposal it is modified the left hand side. 
Indeed, in order to take into account of all the observational evidences: galactic rotation curves, structure formation in the universe, CMB spectrum, bullet cluster, and gravitational lensing, it seems needed to somehow modify  the Einstein's field equations. 
However, in this paper we propose the following different approach, namely: ``understand gravity instead of modifying it''. 

In this document we do not pretend to provide a definitive answer to the ``mystery of missing mass'' or ``missing gravity in the universe'', but we only focus on the galactic rotation curves. Nevertheless, we believe our result to be quite astonishing on both the theoretical and observational side.


The analysis here reported, which follows the previous paper \cite{Li:2019ksm}\footnote{The previous seminal paper \cite{Li:2019ksm} did not address the issue of conformally coupled matter that completely changes the geometrical interpretation of our proposal underlining the crucial role of the asymptotic but harmless spacetime singularity. 
Notice that in \cite{Li:2019ksm} the massive particles break explicitly the conformal invariance, even if slightly, making the solution no longer exact. 
Moreover, we will show in this paper that in presence of conformally coupled matter we do not need to resort to the global structure of space-time and to invoke the small inhomogeneities on the cosmological scale or the presence of the cosmological constant, which will turn out to be too small to affect on the rotation curves on a galactic scale: ``everything will be limited to the single galaxies''.}, is universal and apply to any conformally invariant theory, nonlocal \cite{Krasnikov,kuzmin,Modesto:2011kw,Modesto:2014lga,Modesto:2017sdr}, or local \cite{LM-Sh, LWQG}, that has the Schwarzschild-metric as an exact \cite{Li:2015bqa} and stable solution \cite{Briscese:2018oyx,Briscese:2021mob,Briscese:2018bny,Briscese:2019rii,Modesto:2021okr,Modesto:2021ief,spallucci}.

However, for the sake of simplicity we will focus on Einstein's conformal gravity, whose  general covariant action functional \cite{Deser:1970hs} reads 
\be
 S=\int \! d^4x \sqrt{-\hat{g}} \, \left(\phi^2\hat{R}+6\hat{g}^{\mu\nu}\partial_\mu\phi \partial_\nu \phi- 2 h  \phi^4 \right)
 \label{CEG}  , 
\ee
which is defined on a pseudo-Riemannian spacetime Manifold $\mathcal{M}$ equipped with a metric tensor field $\hat{g}_{\mu\nu}$, a scalar field $\phi$ (the dilaton), and it is invariant under the following Weyl conformal transformation:
\be 
\hat{g}^\prime_{\mu\nu}=\Omega^2\hat{g}_{\mu\nu} \, ,  \quad \phi^\prime=\Omega^{-1}\phi \, ,
\label{ConfInv}
\ee
where $\Omega(x)$ is a general local function. 
In (\ref{CEG}) $h$ is a dimensionless constant that has to be selected extremely small in order to have a cosmological constant compatible with the observed value. However, we here assume $h=0$ because the presence of a tiny cosmological constant will not affect our result (see section (\ref{effectCosmoConstant}) for more details).  
For completeness and in order to show the exactness of the solutions that we will expand later on, we here remind the equations of motion for the theory (\ref{CEG}) for $h=0$, 
\be
&& \phi^2 \hat{G}_{\mu\nu}  =
   \nabla_\nu \partial_\mu \phi^2 - \hat{g}_{\mu\nu} \hat{\Box} \phi^2 
 -  6 \left( \partial_\mu \phi \partial_\nu \phi - \frac{1}{2} \hat{g}_{\mu\nu} g^{\alpha \beta}
 \partial_\alpha \phi \partial_\beta \phi \right) \, ,  \nonumber \\
 &&
 \hat{\Box} \phi = \frac{1}{6} \hat{R} \phi \, . 
 \label{ECGEoM}
 \ee

The Einstein-Hilbert action for gravity is recovered whether the Weyl conformal invariance is broken spontaneously in exact analogy with the Higgs mechanism in the standard model of particle physics (for more details we refer the reader to \cite{Bambi:2016wdn}.)
One possible vacuum of the theory (\ref{CEG}) (exact solution of the equations of motion (\ref{ECGEoM})) is 
$\phi = {\rm const.} = \kappa^{-1}_4 = 1/\sqrt{16 \pi G}$, together with the metric satisfying $R_{\mu\nu} \propto \hat{g}_{\mu\nu}$. 
Therefore, replacing $\phi = 1/\sqrt{16 \pi G} + \varphi$ in the action (\ref{CEG}) and using the conformal invariance to eliminate the gauge dependent Goldstone degree of freedom $\varphi$
, we 
finally end up with the Einstein-Hilbert action in presence of the cosmological constant,
\be
S_{\rm EH} = \frac{1}{16 \pi G} \int d^4 x \sqrt{-\hat{g}} \, \left(\hat{R} - 2 \Lambda \right) ,
\label{EH}
\ee
where $\Lambda$ is consistent with the observed value for a proper choice of the dimensionless parameter $h$ in the action (\ref{CEG}). 
 Ergo, Einstein's gravity is simply the theory (\ref{CEG}) in the spontaneously broken phase of Weyl conformal invariance \cite{NK,Bambi:2016wdn}. 

Let us now expand about the exact solutions in conformal gravity. Given the conformal invariance (\ref{ConfInv}), any rescaling of the metric $\hat{g}_{\mu\nu}$ accompanied by a non trivial profile for the dilaton field $\phi$, is also an exact solution, namely 
 \be
 \hat{g}_{\mu\nu}^*=Q^2(x) \, \hat{g}_{\mu\nu} \, \quad \phi^* = Q(x)^{-1} \, \phi , 
 \label{rescaling}
 \ee
 solve the EoM obtained varying the action (\ref{CEG}) respect to $\hat{g}_{\mu\nu}$ and $\phi$. 
 
 So far the rescaling (\ref{rescaling}) has been used to show how the singularity issue disappearance in conformal gravity
 \cite{NK, Bambi:2016wdn, Bambi:2017yoz, Chakrabarty:2017ysw}.
 However, and contrary to the previous papers, we here focus on a {\em not-asymptotically flat} rescaling of the Schwarzschild metric as a workaround to the non-Newtonian galactic rotation curves. Moreover, the logic in this project is literally opposite to the one implemented in the past works and it is somehow anti-intuitive. In fact, here, instead of removing the spacetime's singularities, we apparently deliberately introduce an unreachable asymptotic singularity. However, as it will be proved later on, the spacetime stays geodetically complete. Indeed, the proper time to reach the singularity at the edge of the Universe will turn out to be infinite.

 Notice that in order to give a physical meaning to the metric (\ref{rescaling}), conformal symmetry has to be broken spontaneously to a particular vacuum specified by the function $Q(x)$. The uniqueness of such rescaling will be discussed in section \ref{Unique}. In the spontaneously broken phase of conformal symmetry observables are still invariant under diffeomorphisms.


\section{The spherically symmetric solution in conformal gravity}
As explained in the introduction, given an exact solution of Einstein's conformal gravity, any rescaled metric is an exact solution too, if the metric is accompanied by a non-trivial profile for the dilaton. Therefore, we here consider the following conformal rescaling of the Schwarzschild spacetime, 
\be
&& d \hat{s}^{ * 2 } = Q^2(x) \left[ -  \left(1-\frac{2GM }{ x}  \right)  dt^2+\frac{dx^2}{1-\frac{2GM}{ x} }+x^2(d\theta^2+\sin^2\theta d\varphi^2) \right] \, , \\
&& Q(x) = \frac{1}{1- \frac{\gamma*}{2} x } \, , \nonumber \\
&& \phi^* = Q(x)^{-1} \, \kappa_4^{-1} \,  ,
\label{Qmetric}
\ee
where we identified $x$ with the radial coordinate. The reason of the particular rescaling $Q(x)$ will be clarify shortly making use of a more suitable radial coordinate. Notice that $Q(x)$ is singular for $x= 2/\gamma^*$, and, therefore, the metric is defined for  $x <  2/\gamma^*$. However, we will prove in the next section that the asymptotic singularity is unattainable, namely it requires an infinte amount of proper time to be reached. As a remnant of the previous work As a remnant of the previous work \cite{Li:2019ksm}, we named $\gamma^*/2$ the free inverse length scale present in the solution. However, in order to manifestly identify the effect of the conformal symmetry it would be useful and more suitable to define: 
$\gamma^*/2 \equiv \ell_{\rm c}$, which we will refer to as the characteristic scale of the system. 

In order to show that the scaling factor $Q(x)$ in (\ref{Qmetric}) is the only one compatible with (i) $g_{00} = -1/g_{11}$  
(we will expand on the uniqueness of the metric in the section \ref{Unique}), we make a coordinate transformation to the usual radial Schwarzschild coordinate ``$r$'', which identifies the physical radius of the two-sphere (ii).
%
The new radial coordinate $r$ is related to $x$ as follows, 
\be
&& x=\frac{r}{1 + \frac{\gamma^*}{2} r } \, , 
\label{xTOr}
\\
&& \frac{ \partial x(r) }{\partial r} = \frac{1}{Q(r)^2} \,  ,
\label{xTOrder}
\ee
and the metric turns into: 
 \be
&& d \hat{s}^{ * 2 }  = -  Q^2(r) \left( 1-\frac{2GM Q(r)}{ r}  \right)  dt^2+\frac{dr^2}{Q^2(r) \left( 1-\frac{2GM Q(r)}{ r}  \right)} + r^2 (d\theta^2+\sin^2\theta d\varphi^2)  \, , 
\nonumber \\
&& Q(r) = 1+ \frac{\gamma^*}{2} r  \qquad \left( \mbox{notice that} \,\, x = \frac{r}{Q(r)} \right) \, , \nonumber \\
&& \phi^*(r) = Q(r)^{-1} \, \kappa_4^{-1}\, .
\label{Qmetricr}
\ee
It deserves to be notice that any rescaling that differs from the one in (\ref{Qmetric}) is not consistent with the two requirements above, namely (i) and (ii). Therefore, in the infinite class of exact solutions conformally equivalent to the Schwarzschild metric, there is only one geometry non-asymptotically flat consistent with $g_{00} = -1/g_{11}$ and two-dimensional transverse area $4 \pi r^2$. Notice that $Q(r)$ in (\ref{Qmetricr}) is only linear in $r$, which is the minimal modification of the metric compatible with analyticity. 
As mentioned above, we will expand further on the uniqueness of the metric in section (\ref{Unique}).

\subsection{Regularity of the Kretschmann and Weyl square invariants}
As a first check of the regularity, we look at the spacetime in $x=2/\gamma^*$.
Since the Schwarzschild spacetime is Ricci flat, before the rescaling the first non-trivial curvature invariant is the Kretschmann scalar, which reads:
\be
     \hat{K} := \hat{R}_{\alpha \beta \gamma \delta} \hat{R}^{\alpha \beta \gamma \delta} = \hat{C}_{\alpha \beta \gamma \delta} \hat{C}^{\alpha \beta \gamma \delta} = \frac{ 48 G^2 M^2}{x^6} \, , 
     \label{Kre}
\ee
where in the last equality we used that $\hat{R}_{\alpha \beta}=0$ and introduced the Weyl tensor $\hat{C}_{\alpha \beta \gamma \delta}$. Under the Weyl rescaling (\ref{ConfInv}) the Weyl tensor, for the following position of the indexes, is invariant, namely 
\be
\hat{C}^{* \alpha}\,_{ \beta \gamma \delta} = \hat{C}^{\alpha}\,_{ \beta \gamma \delta}\, .
\ee
Hence, the Kretschmann scalar (\ref{Kre}) for the metric (\ref{Qmetric}) turns into:
\be
 \hat{C}^{*2} = \hat{C}^{* \alpha}\,_{\beta \gamma \delta} \, \hat{C}^{* \mu}\,_{\nu\rho\sigma} \,  \hat{g}^*_{\alpha \mu} \hat{g}^{* \beta \nu} \hat{g}^{* \gamma \rho} \hat{g}^{* \delta \sigma} 
 = \hat{C}^{\alpha}\,_{\beta \gamma \delta} \, \hat{C}^{\mu}\,_{\nu\rho\sigma} \,  \hat{g}_{\alpha \mu} \hat{g}^{\beta \nu} \hat{g}^{\gamma \rho} \hat{g}^{\delta \sigma} Q^2(x) \, Q^{-2}(x) \, Q^{-2}(x)\,  Q^{-2}(x) = \frac{\hat{C}^2}{Q^4(x)} \, . 
 \label{C}
\ee
Finally, for the metric (\ref{Qmetric}) we find:
\be
\hat{C}^{*2} = \frac{\hat{K}}{Q^4(x)} = \frac{ 48 G^2 M^2}{x^6} \, \left( 1- \frac{\gamma^*}{2} x \right)^4 
\, , 
\ee
which is zero in the limit $x \rightarrow  \gamma^*/2$. The letter point, as we will show explicitly in the next subsection, represents the spatial infinity for the metric (\ref{Qmetric}) because nothing can reach such point in finite proper time. Therefore, the curvature invariant approaches asymptotically zero. 

Using the radial coordinate $r$ the curvature invariant $\hat{C}^{*2}$ turns into ($x = r/Q(r)$):
 \be
\hat{C}^{*2}(r) : = \hat{C}^{*2}(x(r)) = \frac{ \hat{C}^{2}  (x(r))}{Q^4(x(r))} = \frac{ 48 G^2 M^2}{r^6} \, Q^6(r) \, Q^{-4}(r) = 
 \frac{ 48 G^2 M^2}{r^6} \,
\left( 1+ \frac{\gamma^*}{2} r \right)^2 \, , 
\ee
which is now zero for $r \rightarrow + \infty$ according to the inverse coordinate transformation from $x$ to $r$, namely 
\be
 r= \frac{x}{ 1 - \frac{\gamma^*}{2} x} \, , 
 \label{rTOx}
\ee
which diverges to infinity for $x \rightarrow  \gamma^*/2$.

%

On the other hand, the Kretschmann scalar for the metric (\ref{Qmetricr}) is:
\be
&& \hat{K}^*=\frac{1}{2r^6} \left[-2\gamma^{*2}GMr^3(16+12\gamma^*r+3\gamma^{*2}r^2)+\gamma^{*2}r^4(16+12\gamma^*r+3\gamma^{*2}r^2) \right. \nonumber\\
&& 
\hspace{1cm}
\left. +4G^2M^2(24+16\gamma^*r+8\gamma^{*2}r^2+4\gamma^{*3}r^3+\gamma^{*4}r^4) \right]  .
\ee
At large distance the Kretschmann invariant for the metric (\ref{Qmetricr}) tends to a constant $\hat{K}^*\rightarrow 3\gamma^{*4}/2$, which means that the metric (\ref{Qmetricr}) descries asymptotically a spacetime of constant curvature. Indeed, at large scales the metric (\ref{Qmetricr}) approaches the Anti de-Sitter spacetime with 
 scalar curvature $R\rightarrow - 3\gamma^{*2}$ in the limit $r\rightarrow+\infty$. 

Therefore, the two curvature invariants computed above, namely $\hat{C}^{*2}$ and $\hat{K}^*$, are asymptotically finite, and $x=2/\gamma^*$ is not a curvature singularity. The letter point, as we will show explicitly in the next subsection, represents the spatial infinity for the metric (\ref{Qmetric}) because nothing can reach such point in finite proper time. 

Although in this paper we are concerned with the spacetime far outside the event horizon (indeed all the probes in the galaxy are stars and not black holes), the reader may worry about the singularity at $x = 0$ or $r = 0$. However, the resolution of singularities has been rigorously dealt with in several previous articles \cite{NK, Bambi:2016wdn} and the results found there can be exported directly to the metric (\ref{Qmetric}). Indeed, it is sufficient to rescale the latter metric as explicitly done in \cite{Bambi:2016wdn}. For completeness, starting from the metric (\ref{Qmetricr}), we here provide an explicit example of geodetically complete spacetime from short to large distances, namely 
 \be
&& d \hat{s}^{ * 2 }  = S(r) \left[ -  Q^2(r) \left( 1-\frac{2GM Q(r)}{r}  \right)  dt^2+\frac{dr^2}{Q^2(r) \left( 1-\frac{2GM Q(r)}{r}  \right)} + r^2 d \Omega^{(2)}  \right] \, , 
\nonumber \\
&& Q(r) = 1+ \frac{\gamma^*}{2} r  \qquad \left( \mbox{notice that} \,\, x = \frac{r}{Q(r)} \right) \, , \nonumber \\
&& \phi^*(r) = S(r)^{-1/2}  Q(r)^{-1} \, \kappa_4^{-1}\, , \nonumber \\
&& S(r) = 1+ \frac{L^4}{r^4} \, ,
\label{QmetricrSF}
\ee
where $L$ is a parameter with dimension of length (for more details and observational constraints on $L$ see \cite{Bambi:2016wdn, FiniteConformal, Bambi:2017yoz}).

\subsection{Geodetic completion: conformally coupled particles}
For the sake of simplicity from now on in the paper we will remove the label ``$^*$" to the metric and to the dilaton field. 
Let us start with a conformally coupled particle whose action reads: 
\be
S_{\rm cp} = - \int \sqrt{ - f^2 \phi^2 \hat{g}_{\mu\nu} d x^\mu d x^\nu} 
=  - \int \sqrt{ - f^2 \phi^2 \hat{g}_{\mu\nu} \frac{d x^\mu}{d \lambda} \frac{d x^\nu}{d \lambda} } 
\, d \lambda \, , 
\label{Scp}
\ee
where $f$ is a positive constant coupling strength, $\lambda$ is the world-line parameter, and $x^\mu(\lambda)$ is the trajectory of the particle\footnote{Notice that assuming the conformal symmetry to be spontaneously broken to 
 $\phi = \kappa_4^{-1}$ and taking the unitary gauge, the action~(\ref{Scp}) turns into the usual one for a particle with mass $m = f \kappa_4^{-1}$ ($f>0$). Different values for $f$ provide different mass scales.}. From (\ref{Scp}), the Lagrangian reads: 
\be
L_{\rm cp} = - \sqrt{ - f^2 \phi^2 \hat{g}_{\mu\nu} \dot{x}^\mu \dot{x}^\nu } \, , 
\label{Lcp}
\ee
and the translation invariance in the time-like coordinate $t$ implies: 
\be
\frac{\partial L_{\rm cp}}{\partial \dot{t} } = - \frac{f^2 \phi^2 \hat{g}_{tt} \dot{t}}{L_{\rm cp}} = {\rm const.} = - E \quad \Longrightarrow \quad \dot{t} = \frac{L_{\rm cp} E }{f^2 \phi^2 \hat{g}_{tt}} . 
\label{ConstE}
\ee
Since we are interested in evaluating the proper time for the particle to reach the singularity of the universe located in $x= 2/\gamma^*$, we choose the proper time gauge, namely $\lambda = \tau$. Therefore, $E$ can be formally interpreted like the energy of the test-particle and
\be
\frac{d \hat{s}^2}{d \tau^2} = - 1 
\quad \Longrightarrow \quad 
\Bigg\{\hat{g}_{\mu\nu} \dot{x}^\mu \dot{x}^\nu = - 1 \quad {\rm and} 
\quad 
L_{\rm cp}= - f \phi 
\quad \Longrightarrow \quad  \dot{t} = - \frac{ E }{f \phi \, \hat{g}_{tt}} \Bigg\} \, .
\label{PTG}
\ee
Replacing $\dot{t}$ from (\ref{PTG}) in 
\be
\hat{g}_{\mu\nu} \dot{x}^\mu \dot{x}^\nu = - 1 \, , 
\ee
and using the solution of the EOM for $\phi$, namely $\phi = Q^{-1} \kappa^{-1}_4$, we end up with the following first order differential equation for $x(\tau)$, 
\be
&& g_{tt} \dot{t}^2 + g_{xx} \dot{x}^2 = -1 \quad \Longrightarrow \quad 
g_{tt} \frac{E^2}{f^2 \phi^2 g_{tt}^2 }  + g_{xx} \dot{x}^2 = -1 
\quad \Longrightarrow \quad 
\frac{E^2}{f^2 \phi^2 }  + g_{xx} g_{tt} \dot{x}^2 = - g_{tt} \, , 
\\
&& Q(x)^4 \dot{x}^2 + Q(x)^2 \left( 1 - \frac{2 G M}{x} \right) - \frac{E^2 \kappa_4^2}{f^2} Q(x)^2 = 0 \, , 
\label{Geox}
\ee
or, introducing the dimensionless parameter $e^2 \equiv \frac{E^2 \kappa_4^2}{f^2}$, 
\be
 Q(x)^2 \dot{x}^2 =  \frac{2 G M}{x} + e^2 -1  \, .
\label{GeoX}
\ee
Since we are interested to investigate the asymptotic completeness of the spacetime for large $x$, we can assume $x \gg 2 G M$ and (\ref{GeoX}) simplifies to
\be
 Q(x)^2 \dot{x}^2 \simeq e^2 -1  \, ,
\label{GeoX2}
\ee
which must be positive because $Q(x)^2 \dot{x}^2$ is surely positive. 
Replacing $Q(x)$ from (\ref{Qmetric}) into (\ref{GeoX2}) we get:
\be
  \frac{|\dot{x}|}{| 1 - \frac{\gamma^*}{2} x |}  \simeq \sqrt{e^2 -1} > 0  \, .
\label{GeoX3}
\ee
We here would like to study a particle moving from smaller to larger values of $x$, then $\dot{x} >0 $, moreover, $x < \gamma^*/2$, therefore (\ref{GeoX3}) simplifies to 
\be
  \frac{\dot{x}}{1 - \frac{\gamma^*}{2} x}  \simeq \sqrt{e^2 -1} > 0  \, , 
\label{GeoX4}
\ee
and the solution is:
\be
\tau = \frac{2}{\gamma^* \, \sqrt{e^2 -1}} \log\left( \frac{ 1 - \frac{\gamma^*}{2} x_0}{1 - \frac{\gamma^*}{2} x} \right) \qquad {\rm or} \qquad 
x(\tau) = \frac{2}{\gamma^*}  \left[ 1-e^{-\frac{\gamma  \tau }{2}} \left( 1 - \frac{\gamma^*}{2} x_0 \right) \right] \, .
\label{SolGeo}
\ee
According to the solution (\ref{SolGeo}), the proper time to reach the edge of the universe located in $x= 2/\gamma^* = \ell_{\rm c}$ is infinity. Moving to the radial coordinate $r$ defined in (\ref{xTOr}), 
\be
\lim_{x \rightarrow \frac{\gamma^*}{2}} r = \lim_{x \rightarrow \frac{\gamma^*}{2}} \frac{x}{ 1 - \frac{\gamma^*}{2} x} = + \infty \, .
\ee
Therefore, a massive particle will reach $r=+ \infty$ in an infinite amount of proper time. 
Indeed, in the coordinate $r$, the radial geodesic equation (\ref{GeoX}) turns into:
\be
&& Q(r)^2 \dot{r}^2 \left(\frac{\partial x[ r(\tau)] }{ \partial r}\right)^2=  \frac{2 G M Q(r) }{r} + e^2 -1  \, ,  
\nonumber  \\
&& \frac{\dot{r}^2}{Q(r)^2} =  \frac{2 G M Q(r) }{r} + e^2 -1  
\quad (\mbox{where we used} \,\, (\ref{xTOr}) \,\, \rm{and} \,\, (\ref{xTOrder})) \, , 
\label{confrar}  \\
&& \frac{\dot{r}}{Q(r)} \simeq \sqrt{G M \gamma^* + e^2 -1} : = {\rm c} 
\quad {\mbox{for} }  \quad  r \gg 2 G M, \quad \dot{r} > 0 \, , \quad {\rm and} \quad   G M \gamma^* + e^2 -1 >0 \, ,\nonumber  \\
&& \tau = \frac{2 \log \left(\frac{1 + \frac{\gamma^*}{2} r}{ 1 + \frac{\gamma^*}{2}  r_0}\right)}{{\rm c} \, \gamma^* }
\quad \Longrightarrow \quad 
r= \frac{2 e^{\frac{ {\rm c} \gamma^*  \tau }{2}}+ \gamma^*  r_0 e^{\frac{ {\rm c} \gamma  \tau }{2}}-2}{\gamma^* } \, .
\ee
So far we found that the proper time for a particle (conformally coupled) to reach the edge of the Universe is infinite in both $x$ and $r$ coordinates, in the former case the boundary is located at the finite value $x= \gamma^*/2$, in the latter case it is located in $r= +\infty$. 
In the next section we will study the geodesic motion of massless particles.

\subsection{Geodetic completion: massless particles}
For massless particles the correct action, which is invariant under reparametrizations of the world line, $p^\prime= f(p)$, is: 
\be
S_{\gamma} = \int \mathcal{L}_{ \gamma } d \lambda = \int e(p)^{-1} \phi^2\hat{g}_{\mu\nu}  \frac{d x^\mu}{d p} \frac{d x^\nu}{d p} d p \, ,
\label{Lm0}
\ee
where $e(p)$ is an auxiliary field that transforms as $e^\prime(p^\prime)^{-1} = e(p)^{-1} (d p^\prime/d p)$ in order to guarantee the invariance of the action. The action (\ref{Lm0}) is not only invariant under general coordinate transformations, but also under the Weyl conformal rescaling (\ref{ConfInv}). 

The variation respect to $e$ gives:
\be
\frac{ \delta S_{\gamma}}{\delta e} = 0 \quad \Longrightarrow \quad  - \int d p  \, \frac{\delta e}{e^2} \phi^2 \, \hat{g}_{\mu\nu} \, \dot{x}^{\mu} \dot{x}^\nu = 0 \quad \Longrightarrow \quad   d \hat{s}^2 =  \hat{g}_{\mu\nu} dx^\mu d x^\nu = 0 \, ,
\label{ds0}
\ee
which is equivalent to say that massless particles travel along the light cone. 

The variation respect to $x^\mu$ gives the geodesic equation in presence of the dilaton field, namely (in the gauge $e(p)={\rm const.}$)
\be
\frac{D^2(g = \phi^2 \hat{g}) x^\lambda}{d p^2} = 
\frac{D^2(\hat{g}) x^\lambda}{d p^2} + 2 \frac{\partial_\mu \phi}{\phi} \frac{ d x^\mu}{d p} \frac{ d x^\lambda }{d p} 
-  \frac{\partial^\lambda \phi }{\phi} \frac{ d x^\mu}{d p} \frac{ d x_\mu }{d p} 
= 0 \, ,
\label{geom0}
\ee
where $D^2(\hat{g})$ is the covariant derivative respect to the metric $\hat{g}_{\mu\nu}$.

However, when we contract equation (\ref{geom0}) with the velocity $d x_\lambda/d p$ and we use $d\hat{s}^2 =0$ obtained in (\ref{ds0}), we get the following on-shell condition, 
\be
\frac{d x_\lambda}{d p} \, \frac{D^2(\hat{g}) x^\lambda}{d p^2} + 2  \frac{d x_\lambda}{d p} \, \frac{\partial_\mu \phi}{\phi} \frac{ d x^\mu}{d p} \frac{ d x^\lambda }{d p} 
-  \frac{d x_\lambda}{d p} \, \frac{\partial^\lambda \phi }{\phi} \frac{ d x^\mu}{d p} \frac{ d x_\mu }{d p} 
= 0 \quad \Longrightarrow \quad \frac{d x_\lambda}{d p} \, \frac{D^2(\hat{g}) x^\lambda}{d p^2} = 0 \, .
\ee
Therefore, the the covariant derivative $\frac{D^2(\hat{g}) x^\lambda}{d p^2}$ must be proportional to the velocity, namely 
\be
\frac{D^2(\hat{g}) x^\lambda}{d p^2} = f \, \frac{d x^\lambda}{d p}  \quad (f = {\rm const.}) 
\label{constre}
\ee
because the velocity is null on the light cone. 
Under a reparametrization of the world line $q = q(p)$ eq.(\ref{constre}) becomes 
\be
\frac{d^2 x^{\lambda} }{dq^2} + \Gamma^\lambda_{\mu\nu}  \frac{d x^\mu }{d p}  \frac{d x^\nu }{d p} 
=  \frac{d x^\lambda}{d p} \left( \frac{dp}{dq} \right) \left(f \frac{dq}{dp} - \frac{d^2 q}{dp^2} \right) .
\label{parame}
\ee
Choosing the dependence of $q$ on $p$ such us to make vanish the right-hand side of (\ref{parame}), we end up we the geodesic equation in the affine parametrization. Hence, we can redefine $q \rightarrow \lambda$ and, finally, we get the affinely parametrized geodesic equation for photons in the metric $\hat{g}_{\mu\nu}$, 
\be
\frac{D^2(\hat{g}) x^\lambda}{d \lambda^2} = 0 \, .
\label{affine}
\ee
We can now investigate the conservations laws based on the symmetries of the metric $\hat{g}_{\mu\nu}$. Let us consider the following scalar,
\be
\hat{\alpha} = \hat{g}_{\mu\nu} v^\mu \frac{d x^\nu}{d \lambda} = 
\hat{g}_{\mu\nu} v^\mu u^\nu 
\, .
\label{alpha}
\ee
where $v^\mu$ is a general vector and $u^\mu$ the four velocity. Taking the derivative of (\ref{alpha}) respect to $\lambda$ and using the geodesic equation (\ref{affine}) we get:
\be
\frac{d}{d \lambda} \hat{\alpha} = \frac{1}{2}  v^\mu \partial_\mu \hat{g}_{\rho \nu} \frac{d x^\rho}{d \lambda} 
\frac{d x^\nu}{d \lambda} + \hat{g}_{\mu\nu} \partial_\rho v^\mu \frac{d x^\nu}{d \lambda} \frac{d x^\rho}{d \lambda} 
= \frac{1}{2} [\mathcal{L}_v \hat{g} ]_{\rho \nu }  \frac{d x^\rho}{d \lambda} \frac{d x^\nu}{d \lambda} \, , 
\ee
where $[\mathcal{L}_v \hat{g}]$ is the Lie derivative of $\hat{g}_{\mu\nu}$ by a vector field $v^\mu$. 
Thus, if $v^\mu$ is a Killing vector field, namely $[\mathcal{L}_v \hat{g}]=0$, $\hat{\alpha}$ is conserved:
\be
\frac{d}{d \lambda} \left[ \hat{g}_{\mu\nu} v^\mu \frac{d x^\nu}{d \lambda} \right] = 0 
\, . 
\label{CL}
\ee
The metric~(\ref{Qmetric}) is time-independent and spherically symmetric (in particular it is invariant under $t \rightarrow t + \delta t$ and $\varphi \rightarrow \varphi + \delta \varphi$). Therefore, we have the following Killing vectors associated with the above symmetries
\be
\xi^{\alpha} = (1, 0, 0, 0) \, , \quad \eta^{\alpha} = (0, 0, 0, 1) \, .
\ee
Since the metric is independent of the $t$- and $\varphi$-coordinates, according to (\ref{alpha}) we can construct the following conserved quantities
\be
&& e = - \xi \cdot u = - \xi^\alpha u^\beta \hat{g}_{\alpha \beta} = - \hat{g}_{ t \beta} u^\beta = - \hat{g}_{ t t } u^t = 
Q^2(x) \left(  1 - \frac{2 M}{x} \right) \frac{d t}{d \lambda} = Q^2(x) \left(  1 - \frac{2 M}{x} \right) \dot{t} \, , 
\label{Ldott} \\ 
&& \ell = \eta \cdot u = \eta^\alpha u^\beta \hat{g}_{\alpha \beta} =  \hat{g}_{ \phi \beta} u^\beta =  \hat{g}_{ \phi \phi } u^\phi = 
Q^2(x) x^2 \sin^2 \theta \, \dot{\varphi} \, ,
\label{Ldotphi}
\ee 
where the null vector 
\be
u^{\alpha} = \frac{d x^\alpha}{d \lambda} 
\ee
satisfies 
\be
u \cdot u = \hat{g}_{\alpha \beta}  \frac{d x^\alpha}{d \lambda}  \frac{d x^\beta}{d \lambda} = 0 \, ,
\label{NullV}
\ee
as a consequence of (\ref{ds0}). 

From~(\ref{NullV}) in the equatorial plane (i.e. $\theta = \pi/2$), we get the following equation
\be
- \left(  1 - \frac{2 G M}{x} \right) \dot{t}^2 + \frac{\dot{x}^2}{\left(  1 - \frac{2 G M}{x} \right)} 
+ x^2  \, \dot{\varphi}^2 = 0 \, .
\label{uu}
\ee
Note that the rescaling of the metric cancels out in the above equation (\ref{uu}) for null geodesics, but $Q^2(x)$ will appear again when the conserved quantities (\ref{Ldott}) and (\ref{Ldotphi}) are taken into account. Let us solve (\ref{Ldott}) for $\dot{t}$ and (\ref{Ldotphi}) for $\dot{\varphi}$ and, afterwards, replace the results in (\ref{uu}). The outcome is: 
\be
-  \frac{e^2}{Q(x)^4 \left(1 - \frac{2GM}{x} \right)} 
+ \frac{\dot{x}^2}{  1 - \frac{2 G M}{x} } 
+ \frac{\ell^2}{ Q(x)^4 x^2 } = 0 \, .
\label{uu2}
\ee
Let us focus on the radial geodesics (i.e. $\ell=0$), which will be sufficient to verify the geodesic completeness. Equation~(\ref{uu2}) simplifies to:
\be
-  \frac{e^2}{Q(x)^4 } 
+\dot{x}^2 =0 \, \quad \Longrightarrow \quad Q^2(x) | \dot{x} | = e \, .
\label{Geom0}
\ee
The above first order differential equation can be easily integrated for a photon traveling towards the boundary $x=2/\gamma^*$, namely for $\dot{x} > 0$. The result of the integration is:
\be
x(\lambda) = \frac{4 \lambda - 2 \gamma^*  \lambda  x_0  +  4 x_0  }{2 \gamma^*  \lambda - \gamma^{* 2} \lambda  - x_0 + 4} \, , 
\ee
where $x_0$ is the initial position from which the photon is emitted, and 
\be
\lim_{\lambda \rightarrow +\infty} x(\lambda) = \frac{2}{\gamma^*} \, . 
\ee
It turns out that photons cannot reach $x=2/\gamma^*$ for any finite value of the affine parameter $\lambda$. 

In the coordinate $r$ the geodesic equation (\ref{Geom0}) turns into:
\be
{Q(x[r])^4 } 
 \left( \frac{\partial x[r(\tau)]}{\partial r} \right)^2 
\dot{r}^2 = e^2
 \, \quad \Longrightarrow \quad | \dot{r} | = e \, ,
\label{Geom2}
\ee
and a massless particle can reach $r= + \infty$ only for $\lambda = \infty$. The above equation (\ref{Geom2}) has been derived in the appendix (\ref{RadialGeor0}) also directly starting from the metric (\ref{Qmetricr}).

\section{Uniqueness of the solution}\label{Unique}

In the first part of this paper the rescaling of the metric $Q(x)$ was chosen compatibly with the relation $g_{00} = -1/g_{11}$, as evident in the coordinate $r$. In this section we would like to provide three fundamental reasons 
to support such choice. 
(i) The first one is related to the null energy condition, which asserts that $p + \rho \geqslant 0$ \cite{Pons:2014oya}. Indeed, in order to preserve the null energy condition we must impose $g_{00} = -1/g_{11}$. 

(ii) The second one is related to the acceleration of the light in the Newtonian regime. Indeed, if the velocity of light has to remain constant in empty space surrounding a point-like mass, then photons should experience zero acceleration \cite{Dadhich:2012pda}. Using the last result in the previous subsection, namely $| \dot{r} | = e$ we get $\ddot{r} = 0$, which is true only if the relation $g_{00} = -1/g_{11}$ for the components of the metric tensor is satisfied. Let us expand on this point. 
For a general spherically symmetric metric, 
\be
ds^2 = - A(r) dt^2 + B(r) dr^2 + r^2 d \Omega^2 , 
\ee
making use again of (\ref{Ldott}), namely
\be
e = A(r) \dot{t} \, , 
\ee
 and $ds^2 = 0$, the radial geodesic equation reads 
 \be
 \hspace{-0.5cm} 
- A(r) \dot{t}^2  + B(r) \dot{r}^2 = 0  \quad \Longrightarrow \quad 
-  \frac{e^2}{A(r) }  + B(r) \dot{r}^2 = 0 
  \quad \Longrightarrow \quad   \dot{r}^2 = \frac{e^2}{A(r)  B(r)} 
   \quad \Longrightarrow \quad 2 \, \dot{r} \, \ddot{r} = \left(\frac{e^2}{A(r)  B(r)}  \right)^{\prime} \dot{r} \, ,
 \ee
 where $^\prime$ means derivative respect to $r$. Finally,
 \be
 \ddot{r} = \left(\frac{e^2}{ 2 A(r)  B(r)}  \right)^{\prime} \, .
 \ee
 Therefore, in order to do not experience acceleration in the radial coordinate we must have: 
 $A(r) B(r) = {\rm const.}$. Notice that here the radial coordinate is not the physical radial distance because the spacetime is not asymptotically flat.  However, according to the Taylor expansion of (\ref{ellrr}) in the Newtonian intermedium regime $\ell_r \approx r$ and the acceleration above vanishes. 


(iii) Last but not least we should consider the impact of the large distance modification of the Schwarzschild metric on the homogeneity and isotropy of the Universe.

Let us start considering the following coordinate transformation from the radial coordinate $r$ to $\rho$, 
\be
&& \rho = \frac{4 r }{2 (1 + \alpha r + \beta r^2)^{1/2} + 2 + \alpha r}  \, , \\
&& \tau =  \int dt R(t) \, ,  
\label{rrho}
\ee
in the following general not asymptotically flat metric,
 \be
 d \hat{s}^{ * 2 } = -   \left( 1 + \alpha r +   \beta r^2 \right)  dt^2+\frac{dr^2}{\left( 1 + \alpha r + \beta r^2 \right) } + r^2 d \Omega^{(2)}  \, .
\label{QmetricrLargerG}
\ee
The above metric (\ref{QmetricrLargerG}) in the new coordinates reads:
\be
 d \hat{s}^{ * 2 } = \frac{1}{R^2(\tau)} 
 \left[ \frac{ 1 - \frac{\alpha^2 \rho^2}{16} + \frac{\beta \rho^2}{4} }{ \left(1 - \frac{\alpha \rho}{4} \right)^2 - \frac{\beta \rho^2}{4} }\right]^2  
 \left\{
 - d \tau^2 
 + \frac{R(\tau)^2}{\left[ 1 - \left(  \frac{\alpha^2}{16} - \frac{\beta}{4} \right) \rho^2 \right]^2 } \left(d \rho^2 + \rho^2 d \Omega^{(2)} \right)  
\right\} \, .
\label{QmetricrLargerG1}
\ee
Now, in a geometry which is both homogeneous and isotropic about all points, any observer can serve as the origin of the radial coordinate $\rho$; thus in his own local rest frame each observer is able to make the above general coordinate transformation using his own particular $\rho$. Moreover, in conformal gravity we can make an overall rescaling of the metric to finally end up with a comoving Robertson-Walker (RW) spacetime written in spatially isotropic coordinates with spatial curvature $K = \beta - \alpha^2/4$,
\be
 d \hat{s}^{ * 2 } = 
F(\tau, \rho)
 \left[
 - d \tau^2 
 + \frac{R(\tau)^2}{ \left(1 + K \rho^2/4 \right)^2 } \left(d \rho^2 + \rho^2 d \Omega^{(2)} \right)  
\right] \, .
\label{QmetricrLargerG2}
\ee
%
%
%
%
For the case of the metric (\ref{Qmetricr}), taking $r \gg 2 G M$ and $GM \gamma^* \ll 1$, 
 \be
 d \hat{s}^{ * 2 }  \approx -   \left( 1 + \gamma^* r +   \frac{\gamma^{* 2}}{4} r^2 \right)  dt^2+\frac{dr^2}{\left( 1 + \gamma^* r +   \frac{\gamma^{* 2}}{4} r^2 \right) } + r^2 d \Omega^{(2)}  \, .
\label{QmetricrLarger}
\ee
 we can identify the constants $\alpha = \gamma^*$ and $\beta = \gamma^{* 2}/{4}$, and in the new coordinates $(\tau, \rho)$ the metric (\ref{QmetricrLarger}) takes the following RW form,
 \be
  d \hat{s}^{ * 2 } = \frac{1}{R^2(\tau)} 
  \frac{ 1 }{ \left(1 -  \frac{\gamma^*}{2} \rho\right)^2 }  
 \left[
 - d \tau^2 
 + R(\tau)^2 \left( d \rho^2 + \rho^2 d \Omega^{(2)} \right)  
\right] \, ,
\label{QmetricrCosmo}
\ee
which coincides with the metric (\ref{Qmetric}) for $x\gg 2 GM$ upon reintroducing the time coordinate $t$ defined in (\ref{rrho}). 

Therefore, the metric proposed in this paper is the {\em only one} that does not affect the homogeneity of the Universe at large scales. Finally, we notice that the metric (\ref{QmetricrLarger}) is asymptotically (for large $r$) Anti-de Sitter, whose stability is guarantee from the fact that it comes form a rescaling of the Schwarzschild metric, which is known to be stable. 

\section{The cosmological constant is not an issue in our model} \label{effectCosmoConstant}
It is commonly accepted that the value of the cosmological constant is non zero ($\Lambda\sim10^{-56}$cm$^{-2}$). 
Therefore, we will have to more correctly consider the rescaling of the Schwarzschild-de Sitter spacetime instead of (\ref{Qmetric}) or (\ref{Qmetricr}), namely 
\be
&& d \hat{s}^{ * 2 }  = Q^2(x) \left[ -  \left(1-\frac{2GM}{c^2 x} - \frac{\Lambda}{3} x^2 \right) dt^2+\frac{dx^2}{1-\frac{2GM}{ x} - \frac{\Lambda}{3} x^2}+x^2 \Omega^{(2)} \right] \, , \\
&& Q(x) = \frac{1}{1- \frac{\gamma*}{2} x } \, ,
\label{QmetricdS}
\ee
or in the radial coordinate $r$, 
 \be
&& d \hat{s}^{ * 2 }  = -  Q^2(r) \left( 1-\frac{2GM Q(r)}{r} - \frac{\Lambda}{3} \frac{r^2}{Q^2(r)} \right) c^2 dt^2+\frac{dr^2}{Q^2(r) \left( 1-\frac{2GM Q(r)}{ r} - \frac{\Lambda}{3} \frac{r^2}{Q^2(r)} \right)} + r^2 d\Omega^{(2)}  \, , 
\nonumber \\
&& Q(r) = 1+ \frac{\gamma*}{2} r   \, .
\label{QmetricrdS}
\ee
Notice that the metric is till in the form $g_{00}(r) = - 1/g_{11}(r)$. 
If we focus on (\ref{QmetricrdS}) and we consider the limit $r \gg 2 G M$ together with the approximation $G M \gamma^* \ll 1$, the metric (\ref{QmetricrdS}) simplifies to:
 \be
 d \hat{s}^{ * 2 }  &\approx&  - \left(Q^2(r)-\frac{\Lambda}{3}r^2\right)dt^2+\frac{dr^2}{\left(Q^2(r)-\frac{\Lambda}{3}r^2\right)}+r^2d\Omega^{(2)} \nonumber \\
&&  = - \left( 1 + \gamma^* r + \frac{\gamma^{* 2} }{4} r^2 - \frac{\Lambda}{3}r^2\right)dt^2+\frac{dr^2}{\left( 1 + \gamma^* r + \frac{\gamma^{* 2}}{4} r^2 -\frac{\Lambda}{3}r^2\right)}+r^2d \Omega^{(2)}
\, . 
\label{QmetricrdSlimit}
\ee
However, since $\gamma^{*2}\gg\Lambda$ (we will see later that $\gamma^* \sim 10^{-21} {\rm m}^{-1}$) then the presence of the cosmological constant will not affect our analysis
\footnote{Looking at the Mannheim's paper \cite{Mannheim:2010xw}, in the appendix A5 the potential is defined as usually like $-(g_{00} + 1)/2$ (see the paragraph before formula A43 and formula A45). However, this is inconsistent with the physical velocity that we get from the metric (\ref{QmetricrLargerG}). Indeed, the usual derivation of the potential, which one can find for example in Landau's book ``Classical Field Theory'', does not work for spacetimes not asymptotically flat, which is the case of (\ref{QmetricrLargerG}). The correctness of Mannheim's paper lies in the fact that the scales in his model are much larger than the galactic extension.
It deserves to be notice that for special values of $\gamma_0$ and $k$ in Mannheim's paper, namely $\gamma_0 = \gamma^*$ and $k^2 = - \gamma^{*2}/4$, the exact solution (\ref{QmetricrLargerG}) (it is (5) in Mannheim's paper) of Weyl conformal gravity turns out to be a conformal rescaling of the Minkowski spacetime. 
}.

Finally we want to make the following speculative comment. 
It deserves to be notice that the value of the radius of de Sitter's spacetime (proportional to the inverse of the square root of the cosmological constant) is about the radius of the Universe. Therefore, for $\ell_c$ comparable to the radius of the Universe the two contributions quadratic in $r$ in (\ref{QmetricrdSlimit}) can cancel each other.


\section{The orbital velocity}
In this section we compute the orbital velocity 
of a conformally coupled probe particle on the equatorial plane in the geometry (\ref{Qmetricr}) and (\ref{Qmetric}) respectively assuming zero radial velocity. 
For completeness let us remind here the Action for a conformally coupled particle (\ref{Scp}),
\be
S_{\rm cp} = - \int \sqrt{ - f^2 \phi^2 \hat{g}_{\mu\nu} d x^\mu d x^\nu} 
=  - \int \sqrt{ - f^2 \phi^2 \hat{g}_{\mu\nu} \frac{d x^\mu}{d \lambda} \frac{d x^\nu}{d \lambda} } 
\, d \lambda \, ,
\label{Scp2}
\ee
from which the Lagrangian reads:
\be
L_{\rm cp} = - \sqrt{ - f^2 \phi^2 \hat{g}_{\mu\nu} \dot{x}^\mu \dot{x}^\nu } \, . 
\label{Lcp2}
\ee
Since both the metrics (\ref{Qmetricr}) and (\ref{Qmetric}) are nvariant whether we make the replacements $t \rightarrow t + {\rm const}.$ and $\varphi \rightarrow \varphi + {\rm const}.$. Therefore, from the Lagrangian (\ref{Lcp2}) we obtain the following conserved quantities (for $\theta=\pi/2$), 
\begin{gather}
\frac{\partial{L_{\rm cp}}}{\partial \dot{t}} = - \frac{ f^2 \phi^2\hat{g}_{tt}\dot{t}}{{L_{\rm cp}}}
 = - E \, , \qquad 
	\frac{\partial{L_{\rm cp}}}{\partial \dot{\varphi}} = - \frac{ f^2 \phi^2\hat{g}_{\varphi\varphi}\dot{\varphi}}{{L_{\rm cp}}}=  \ell \,  .
	\label{conserva}
\end{gather}
In the proper time gauge $\lambda \equiv \tau$, $d \hat{s}^2/d \lambda^2 =-1$ and $L_{\rm cp} = - f \phi$. Hence, from 
(\ref{conserva}), 
\be
\dot{t}= \frac{E}{f \, \phi \, \hat{g}_{tt}} \, , \qquad 
	\dot{\varphi} = - \frac{\ell}{f \, \phi \, \hat{g}_{\varphi\varphi}} \, .
	\label{dotti}
\ee

\subsubsection{The orbital velocity in the metric (\ref{Qmetricr})}
 Let us in this section focus on the metric (\ref{Qmetricr}). Again in the proper time gauge and for $\theta = \pi/2$,
 \be
 \frac{d \hat{s}^2}{d \lambda^2} = - 1 \quad \Longrightarrow \quad \hat{g}_{tt} \dot{t}^2 + \hat{g}_{rr} \dot{r}^2 + \hat{g}_{\varphi \varphi} \dot{\varphi}^2 = -1 \, ,
 \label{orbitalprop}
 \ee
 and replacing (\ref{dotti}) in (\ref{orbitalprop}), we get:
 \be
 \hat{g}_{tt} \left( \frac{E}{f \, \phi \, \hat{g}_{tt}} \right)^2 + \hat{g}_{rr} \dot{r}^2 
 + \hat{g}_{\varphi \varphi} \left(  - \frac{\ell}{f \, \phi \, \hat{g}_{\varphi\varphi}}  \right)^2 = -1 
 \quad \Longrightarrow \quad
    \frac{E^2}{f^2 \, \phi^2 \, \hat{g}_{tt}}  + \hat{g}_{rr} \dot{r}^2 
 +  \frac{\ell^2}{f^2 \, \phi^2 \, \hat{g}_{\varphi\varphi}}   = -1 \, .
 \ee
Since we are interested in the orbital motion we can take $\dot{r} = 0$ and we end up with the following constraint equation, 
 \be
    \frac{E^2}{f^2 \, \phi^2 \, \hat{g}_{tt}}  +  \frac{\ell^2}{f^2 \, \phi^2 \, \hat{g}_{\varphi\varphi}}   = -1 
    \quad \Longrightarrow \quad 
   - \frac{E^2}{f^2 \, Q^{-2}(r) \kappa_4^{-2} \, Q^2(r) \left( 1 - \frac{2 G M Q(r) }{r} \right)}  +  \frac{\ell^2}{f^2 \, Q^{-2}(r) \kappa_4^{-2}  \, r^2}   = -1 
    \, .
    \label{constraint}
 \ee
In order to extract a simple relation for the ratio between $\ell^2$ and $E^2$, we take the derivative of equation (\ref{constraint}) respect to $r$, 
\be
E^2 \frac{\frac{d}{d r} \left(  1 - \frac{ 2 G M Q(r) }{r} \right)}{\left(  1 - \frac{ 2 G M Q(r) }{r} \right)^2} + 
\ell^2 \frac{d}{d r} \left( \frac{Q^2(r)}{r^2} \right) = 0
\quad \Longrightarrow \quad \frac{\ell^2}{E^2} = \frac{2 G M r^3}{(2 + r \gamma^* ) (r-G M (2+r \gamma^* ))^2} 
\, . 
\label{LE}
\ee

The physical velocity on the equatorial plane and along the $\varphi$-direction reads:
\begin{equation}
v=\frac{\sqrt{\hat{g}_{\varphi\varphi}}}{\sqrt{-\hat{g}_{tt}}}
\,
\frac{d\varphi}{dt}=\frac{\sqrt{\hat{g}_{\varphi\varphi}}}{\sqrt{-\hat{g}_{tt}}}
\, 
\frac{\dot{\varphi}}{\dot{t}} \, , 
\label{vr}
\end{equation}
where dot stays for the derivative respect to the proper time $\tau$. Replacing $\dot{t}$ and $\dot{\varphi}$ in (\ref{dotti}) into (\ref{vr}) we get:
\begin{equation}
	v^2=-\frac{\hat{g}_{tt}}{\hat{g}_{\varphi\varphi}}\frac{\ell^2}{E^2} \, , 
	\label{v2gene}
\end{equation}
where we finally replace (\ref{LE}),
\begin{equation}
	v^2= \frac{GM(2+\gamma^*r)}{2( r- GM (2+ \gamma^*r) )} = \frac{G M Q(r)}{r - 2 G M Q(r) }
		\, .
\end{equation}
In the limit of $r\gg2GM$, namely far from the Schwarzschild radius, the velocity turns into:
\be
v^2= \frac{G M Q(r)}{r(1 - G M \gamma^*) } \, , 
\label{v2r2GM}
\ee
and if we also assume $GM\gamma^*\ll1$, 
\begin{equation}
	v^2= \frac{ G M Q(r)}{r} = \frac{GM}{r}+\frac{GM\gamma^*}{2}
	\label{v2r}
\end{equation}
which asymptotically approaches the constant value:
\be
v^2 \, \rightarrow \, v^2_{\infty} =  
\frac{GM\gamma^*}{2}. 
\label{vinfinity}
\ee
Let us now express the velocity in terms of the physical length $\ell_r$ in place of the radial coordinate $r$. What we need is the physical radial length, namely
\be
\ell_r &= & \int \sqrt{ \hat{g}_{rr} } \,  dr + {\rm const} 
=  \int \frac{ dr }{Q(r) \sqrt{1 - \frac{2 G M Q(r)}{r}}} + {\rm const}
\approx  \int \frac{ dr }{ \sqrt{1 - G M \gamma^* } \, (1+ \frac{\gamma^*}{2} r)} + {\rm const}
\nonumber \\
& = & \frac{2 \log (2 + \gamma^* \, r)}{\gamma^*  \sqrt{1 -  G M \gamma^*}} + {\rm const}
\, .
\ee
where in the last by one step we have integrated for $r \gg 2 G M$. 
Finally, we fix the integration constant imposing that $\ell_r(r=0) = 0$, 
\be
\ell_r = \frac{2 \log\left(1 + \frac{\gamma^*}{2} \, r \right)}{\gamma^*  \sqrt{1 -  G M \gamma^*}} \, . 
\label{ellrr}
\ee
Notice that in the intermedium Newtonian regime, namely $r \ll 2/\gamma^*$, and for $GM\gamma^*\ll1$, $\ell_r \approx r$.
The inverse relation $r(\ell_r)$ reads:
\be
r(\ell_r)= \frac{2 \left(e^{\frac{1}{2} \gamma^*  \ell_r \sqrt{1 - G M \gamma^*}}-1\right)}{\gamma^* } 
\approx \frac{2 \left(e^{\frac{1}{2} \gamma^*  \ell_r}-1\right)}{\gamma^* } 
\, ,
\label{rlr}
\ee
where the last approximation comes again from $GM\gamma^*\ll1$ 
(notice that also $r(\ell_r = 0) = 0$). 

Replacing (\ref{rlr}) in (\ref{v2r2GM}), we get the physical velocity square, namely 
\be
v^2(\ell_r) = \frac{G M \gamma^*}{4} \frac{\left[ 1 + \coth \left( \frac{1}{4} \ell_r \gamma^* \sqrt{1 - G M \gamma^*} \right)\right] }{1 -  G M \gamma^*} \,  ,
\label{v2lr}
\ee
which further simplifies for $GM\gamma^*\ll1$, 
\be
\boxed{v^2(\ell_r) = \frac{G M \gamma^*}{4} \left[ 1 + \coth \left( \frac{ \ell_r \gamma^*}{4} \right)\right] } \,  .
\label{v2lr2}
\ee
The above astonishing simple analytic result correctly interpolates between the Newtonian's velocity and the asymptotic constant value (\ref{vinfinity}). It deserve to be notice that for small $\gamma^*$, namely $\ell_c \gg r_g = 2 G M$ ($r_g$ is the Schwarzschild radius), the exact result (\ref{v2lr}) and the velocity (\ref{v2r}) are extremely close each other. Therefore, the following replacement is a good approximation of (\ref{v2r}),
\be
v^2(\ell_r) = \frac{GM}{\ell_r}+\frac{GM\gamma^*}{2} \,  .
\label{v2rlr}
\ee

\subsubsection{The orbital velocity in the metric (\ref{Qmetric})}
In this section we compute again the velocity square, but now for the metric (\ref{Qmetric}). This computation not only will provide a further check of our result (\ref{v2lr2}), but also will make more explicit the crucial role of the asymptotic singularity in $x=2/\gamma^* = \ell_{\rm c}$. 

According to the previous section ($\dot{t}$ and $\dot{\varphi}$)(\ref{dotti}), the velocity (\ref{vr}), and the velocity square (\ref{v2gene}) are general and independent on the metric. However, the ratio $\ell^2/E^2$ it does depend on the metric. Indeed, the proper time gauge for the metric (\ref{Qmetric}) reads:
\be
 \frac{d \hat{s}^2}{d \lambda^2} = - 1 \quad \Longrightarrow \quad \hat{g}_{tt} \dot{t}^2 + \hat{g}_{xx} \dot{x}^2 + \hat{g}_{\varphi \varphi} \dot{\varphi}^2 = -1 \,
 \label{orbitalpropx}
 \ee
which, for $x =$const. and replacing the metric (\ref{Qmetric}) within, turns into:
 \be
&& \hat{g}_{tt} \left( \frac{E}{f \, \phi \, \hat{g}_{tt}} \right)^2 
 + \hat{g}_{\varphi \varphi} \left(  - \frac{\ell}{f \, \phi \, \hat{g}_{\varphi\varphi}}  \right)^2 = -1 
 \quad \Longrightarrow \quad
    \frac{E^2}{f^2 \, \phi^2 \, \hat{g}_{tt}}  
 +  \frac{\ell^2}{f^2 \, \phi^2 \, \hat{g}_{\varphi\varphi}}   = -1 
 \nonumber \\
 &&  \Longrightarrow \quad 
  \frac{E^2}{f^2 \, Q^{-2}(x) \, Q^2(x) \left( 1 - \frac{2 G M}{x} \right)  }
 +  \frac{\ell^2}{f^2 \,  Q^{-2}(x) \, Q^2(x)  x^2}   = -1 
 \quad  \Longrightarrow \quad 
  \frac{E^2}{f^2 \,  \left( 1 - \frac{2 G M}{x} \right)  }
 +  \frac{\ell^2}{f^2 \,   x^2}   = -1 \, ,
 \label{constraint0}
 \ee
which is independent on the rescaling $Q(x)$. Taking the derivative of (\ref{constraint0}) respect to radial coordinate $x$, we find:
\be
\frac{\ell^2}{E^2} = - \frac{g_{tt}^\prime}{g_{\varphi \varphi}^\prime } \, \frac{g_{\varphi \varphi}^2}{g_{tt}^2} \, .
\label{lex}
\ee
where we defined:
\be
g_{tt} = - \left( 1 - \frac{2 G M}{x} \right) \, , \quad g_{ \varphi \varphi} = x^2 \, .
\label{Sch}
\ee
The one above is not just a definition, but the Schwarzschild metric before to introduce the rescaling $Q(x)$.
Substitution of (\ref{lex}) in the velocity square (\ref{v2gene}) and making use of (\ref{Sch}) together imply:
\be
v^2(x) = \frac{g_{tt}^\prime}{g_{\varphi \varphi}^\prime }\frac{g_{\varphi \varphi}}{g_{tt}} = \frac{G M}{x - 2 GM} \approx \frac{GM}{x} \, , 
\label{v2x}
\ee
where in the last equality we assumed $x \gg 2 G M$. 

The result just found for the velocity square may seem trivial and obvious, but it is actually rich in geometric meaning. Indeed, it is exactly the Newtonian's result in the radial coordinate $x$. However, we must remind that the larger value for $x$ is $2/\gamma^*$ and, therefore, the minimum asymptotic value for the velocity square is $GM\gamma^*/2$ in perfect agreement with (\ref{vinfinity}). This is clearly due solely to the singular structure of the conformal geometry in the unattainable asymptotic point $x = 2/\gamma^*$.

In order to complete the section we now express the velocity square in terms of the physical length $\ell_x$ that we set about calculating, 
\be
\ell_x = \int \sqrt{g_{xx}} \, dx + {\rm const.} = \int dx \frac{  Q(x) }{ \sqrt{ 1 - \frac{2 G M}{x} } }+ {\rm const.} \approx 
\int Q(x) \, dx + {\rm const.} 
= - \frac{2}{\gamma^*} \log \left( 1 - \frac{\gamma^*  x}{2} \right) 
\, , 
\label{lx}
\ee
where again we assumed $x \gg 2 G M$ and we fixed the integration constant imposing $\ell_x(x=0)  = 0$. 
Notice that $\ell_x \rightarrow +\infty$ for $x\rightarrow 2/\gamma^*$. 

It is straightforward to invert (\ref{lx}), 
\be
x(\ell_x) =  \frac{2}{\gamma^*} \left(1-e^{- \frac{\ell_x \gamma^*}{2}}  \right) \, .
\ee
Replacing the above expression in the velocity square (\ref{v2x}) we find:
\be
v^2(\ell_x) = \frac{G M}{x(\ell_x)} = \frac{G M \gamma^*}{4}  \left[1 + \coth \left( \frac{\ell_x \gamma^*}{4} \right) \right]\, , 
\label{v2lx2}
\ee
which of course agrees with (\ref{v2lr2}), which is also expressed in terms of the physical distance. Notice that $\ell_x \equiv \ell_r$ because there is only one physical observable distance in nature. 

\section{Newtonian effective theory and Gravitational Potential}
In order to derive the effective gravitational potential we start from the orbital velocity in terms of the physical distance. Indeed, in Newtonian physics we only deal with physical lengths and the Lagrangian simply reads:
\be
\mathcal{L}_{\rm N} = \frac{1}{2} m \left( \frac{ d \vec{r}}{d t} \right)^2 - m \Phi(|\vec{r}|)
= \frac{1}{2} m \left(\dot{\ell}_r^2 + \ell_r^2 \dot{\varphi}^2\right)  - m \Phi( \ell_r) 
 \, , 
\ee
where $|\vec{r}| = \ell_r$, $m$ is the mass of a probe particle, and we assumed to be on the equatorial plane $\theta = \pi/2$. 
From the Lagrangian above the EoM, assuming $\dot{\ell}_r = 0$, is:
\be
\ell_r  \dot{\varphi}^2 = 
\frac{v^2(\ell_r)}{\ell_r} = \frac{ \partial \Phi(\ell_r)}{\partial \ell_r} = - E_{r}(\ell_r) 
\quad \Longrightarrow \quad 
v^2(\ell_r) =  - \ell_r \, E_{r}(\ell_r) 
\, ,
\label{v2N}
\ee
where for future reference we also defined the gravitational field $\vec{E} = - \vec{\nabla} \Phi$. 

Therefore, the effective potential can be obtained simply integrating (\ref{v2lr2}) or (\ref{v2lx2}),
\be
\Phi(\ell_r)= \int_{+ \infty}^{\ell_r } d \ell_r^\prime \frac{v^2(\ell_r^\prime)}{\ell_r^\prime} \, .
\label{intv23}
\ee
However, the velocity in (\ref{intv23}) can be very well approximated making use of (\ref{v2rlr}), and the integral 
(\ref{intv23}) can be easily computed to give the following result, 
\be
\Phi(\ell_r) \approx  - \frac{GM}{\ell_r} + \frac{GM\gamma^*}{2} \log ( \ell_r ) + {\rm const.} \, . 
\ee

Now we have to consider the contribution of all the stars in a galaxy gravitationally acting on a probe star. This consists on integrating the potential in cylindrical coordinates after having introduced the following three vectors: $\vec{R}$, which points from the center of the galaxy to the probe star, $\vec{R}^\prime$ from the center of the galaxy to one of its stars, and $\vec{r}$ pointing from a star in the galaxy to the probe star. Therefore, we have: $\vec{r} = \vec{R} - \vec{R}^\prime$ and the contribution to the potential due to any star in the galaxy is:
\be
\Phi(| \vec{R} - \vec{R}^\prime |) = - \frac{GM}{| \vec{R} - \vec{R}^\prime |} + \frac{GM\gamma^*}{2} \log (| \vec{R} - \vec{R}^\prime | ) + {\rm const.} 
 \equiv \Phi_0 + \Phi^{\rm log}
\, .
\label{singcon}
\ee
Notice that we replaced $\ell_r$ with $| \vec{R} - \vec{R}^\prime |$ because the Newtonian effective theory is defined in flat spacetime.  

Let us now consider a thin disk galaxy model with exponential  distribution of matter that decays at large distance.  We assume that the mass of each star is $M_{\odot}$(solar mass) and the distribution of stars is described as follows (in cylindrical coordinates: $R, \varphi, z$):
\be\label{rho}
\rho(R^\prime, z^\prime) = \Sigma_0 \, e^{ - \frac{R^\prime}{R_0}} \, \delta(z^\prime) 
\, , \quad [\rho] = L^{-3} \, , 
\ee
where $z$ is the coordinate orthogonal to the galaxy plane. Moreover, $R_0$ is the radius of the galaxy and $\Sigma_0$ is related to the number of stars of mass comparable to the solar mass in the galaxy, i.e. 
\be
N^* & = & \int_0^{+ \infty} d R^\prime \, R^ \prime  \int_0^{2 \pi} d \varphi^\prime \int_{- \infty}^{+ \infty} d z^\prime
\, \rho(R^\prime, z^\prime) \nonumber \\
&=& 
\int_0^{+ \infty} d R^\prime \, R^ \prime \int_0^{2 \pi} d \varphi^\prime \int_{- \infty}^{+ \infty} d z^\prime
\Sigma_0 \, e^{ - \frac{R^\prime}{R_0}} \, \delta(z^\prime) = 2 \pi \Sigma_0 R_0^2 \quad \Longrightarrow \quad 
\Sigma_0 = \frac{ N^*}{2\pi R_0^2} \,.
\label{sigma0}
\ee
In order to get the total contribution to the gravitational potential we have to integrate over all the star in the galaxy each of them of solar mass $M_{\odot}$, namely:
\be
&& \hspace{-0.5cm}
\Phi_{\rm T}(R, z) = \int_0^{+ \infty} d R^\prime \, R^ \prime \int_0^{2 \pi} d \varphi^\prime \int_{- \infty}^{+ \infty} d z^\prime
\, \rho(R^\prime, z^\prime) 
\, \Phi(R, R^\prime, z, z^\prime) \, ,\\
&& \hspace{-0.5cm}
\Phi(R, R^\prime, z, z^\prime) = - \frac{ G M_\odot}{\left[ R^2 + R^{\prime 2} - 2 R \, R^\prime \cos \varphi^\prime + (z - z^\prime)^2 \right]^{\frac{1}{2}} }
+ \frac{ G M_\odot \gamma^*}{2} \log \left\{ \! \frac{
\left[ R^2 + R^{\prime 2} - 2 R \, R^\prime \cos \varphi^\prime + (z - z^\prime)^2 \right]^{\frac{1}{2}}}{\ell}
\! \right\}, \nonumber 
\ee
where $R$ is the distance of the probe star from the galactic center in cylindrical coordinates and $R_0$ is the characteristic scale of the galaxy. Since $\Phi(R,R',z,z')$ consists of two parts, we will integrate the two contributions of the potential separately obtaining the two corresponding contributions to the velocity square. Finally, $\ell$ is the scale coming from the integration constant that can not be zero since the potential grows with the distance. However, we do not have to worry about such scale because it will disappear in the orbital velocity that is related to the force and not the potential.

For the Newtonian potential contribution to (\ref{singcon}), namely $\Phi_0=-GM/|\vec{R}-\vec{R'}|$, and assuming the density profile (\ref{rho}), the rotation velocity square of a probe star was computed in \cite{Mannheim:2010xw} and the result is:
\begin{equation}
	v^2_0=\frac{GN^*M_{\odot}R^2}{2R_0^3} \left[ I_0 \left( \frac{R}{2R_0} \right)K_0 \left( \frac{R}{2R_0} \right) -I_1\left( \frac{R}{2R_0} \right) K_1 \left( \frac{R}{2R_0} \right) \right] \, , 
	\label{v2M}
\end{equation}
where $I_0$, $I_1$ are the modified Bessel functions of first kind and $K_0, K_1$ are the modified Bessel functions of second kind. In (\ref{v2M}) $M = N^*M_{\odot}$ is the mass of the all stars in the galaxy. 

In order to compute the logarithmic contribution to the potential in (\ref{singcon}), namely 
\be
\Phi^{\log} = \frac{GM\gamma^*}{2} \log \frac{|\vec{R}-\vec{R'}|}{\ell} \, ,
\label{Phi0}
\ee
we can use the Gaussian theorem 
\be
{\rm div} \vec{E} = - 4 \pi G \rho \, , 
\label{Gauss}
\ee
to sum over all the stars in the galactic disc. 
Notice that we can assume the sources of the logarithmic potential to be wires because of the cylindrical symmetry of the galaxy. 

Due to the above logarithmic correction (\ref{Phi0}), the gravitational field in cylindrical coordinates (we here fix the origin in $\vec{R}' = 0$) is attractive and reads: 
\be
E_R = - \frac{\partial \Phi^{\log} (R)}{\partial R} = - \frac{GM\gamma^*}{2 R} . 
\label{Eg}
\ee
Integrating (\ref{Gauss}) on a three-dimensional volume $V$ with boundary $\partial V$ in Cylindrical coordinates, we can infer about the energy density $\rho_{\rm s}( \vec{x} ) = \rho_0 \delta(x) \delta(y)$ of a single wire-like source, namely  
\be
&&\int_V {\rm div} \vec{E} \, dv = {\rm Flux}\Big|_{\partial V} (E) = - 4 \pi G \int_V dv \rho_{\rm s}( \vec{x} ) \, , \nonumber \\
&&
2\pi R\Delta z\,   E_R  = - 4 \pi  G \rho_0 \, \Delta z \, .
\label{IntGauss}
\ee
Replacing (\ref{Eg}) in (\ref{IntGauss}) we finally find $\rho_0$,
\be
 \rho_0 = - \frac{R \, E_R}{2} = \frac{M \gamma^*}{4}, 
\ee
and the potential can be recast in the following form in terms of the energy density, 
\be
\Phi^{\log} = 2 G \rho_0 \, \log \frac{|\vec{R}-\vec{R'}|}{\ell} \, .
\label{Phi0rho}
\ee

If the gravitational sources and the probe star are all located in the same plane (we here assume the galactic disk to be in $z=0$ in cylindrical coordinates), then $\Phi^{\log}$ is analogous to the Newtonian potential of $N^*$ massive infinite wires each with uniform density $\rho_0$ and generating a logarithmic gravitational potential. 
%

Assuming the principle of linearity of the gravitational forse and then of the gravitational potential, we can now apply again the Gauss' theorem to all the stars in the galaxy that are descried by the energy density profile in cylindrical coordinates:
\be
\rho_{\rm s}( \vec{x} ) = \rho_0 \delta(x) \delta(y) \qquad \longrightarrow \qquad 
\rho_{N^*}(R) = \rho_0 \, \Sigma_0 \, e^{- \frac{R}{R_0}}  =
\frac{M_{\odot}\gamma^*}{4} \Sigma_0 \, e^{- \frac{R}{R_0}}  \, , \quad [\rho_{N^*}] = M L^{-3} \, ,
\label{rhow}
\ee
where we assumed any star to have mass $M_{\odot}$. 
Notice that (\ref{rhow}) is an energy density while (\ref{rho}) is a density distribution. 

Finally, the Gaussian theorem making use of the above energy density (\ref{rhow}) gives:
\be
- 2 \pi R\Delta z \, E^{\rm T}_R(R)  = 4\pi G\int_0^R\int_0^{2\pi}  R' dR' d\varphi  \, \rho_{N^*}(R')\int_0^{\Delta z}dz\quad \Longrightarrow \quad 
E_R^{\rm T} (R) = - \frac{4\pi G}{R}\int_0^R \rho_{N^*}(R') R' dR' \, .
\label{TotV}
\ee
%
%
Using (\ref{v2N}) and upon integration of (\ref{TotV}), the contribution to the rotation velocity square (\ref{v2N}) due to the logarithmic term in the potential reads,
\be
v^{2}_{\rm log} & = & - E_R^{\rm T} (R) R = \pi GM_{\odot}\gamma^*\Sigma_0\int_0^Re^{- \frac{R^\prime}{R_0} } R^\prime dR^\prime =\frac{GM_{\odot}\gamma^*}{2} \, 2\pi\Sigma_0R_0^2 \, \left[ 1- \left( 1+\frac{R}{R_0} \right)e^{- \frac{R}{R_0} } \right]\nonumber \\
&=&\frac{GN^*M_{\odot}\gamma^*}{2} \left[ 1- \left( 1+\frac{R}{R_0} \right)e^{- \frac{R}{R_0} } \right].
\label{Dv2}
\ee
Finally, taking the sum of (\ref{v2M}) and (\ref{Dv2}) the total contribution to the velocity square reads:
\be
\hspace{-0.4cm}
\label{vtot}
\boxed{v^2(R) =
\frac{GN^*M_{\odot}R^2}{2R_0^3} \left[ I_0 \left( \frac{R}{2R_0} \right)K_0 \left( \frac{R}{2R_0} \right) -I_1\left( \frac{R}{2R_0} \right) K_1 \left( \frac{R}{2R_0} \right) \right]
+\frac{GN^*M_{\odot}\gamma^*}{2} \left[ 1- \left( 1+\frac{R}{R_0} \right)e^{- \frac{R}{R_0}} \right] } 
\ee
which is constant for large $R$, namely 
\be
v^2(R) \,\, \rightarrow \,\, \frac{G N^*M_{\odot} \gamma^*}{2}   \quad {\rm for} \quad R \rightarrow + \infty. 
\ee

\section{The Tully-Fisher relation} 


As we have said several times, in conformal gravity we are free to rescale the metric by an overall factor that will depend on at least one undetermined length scale. In our model the length scale is $\ell_c = 2/\gamma^*$, which turns out to be of the same order of magnitude of the galaxy (see next section).
However, if we focus our attention on a single star in the galaxy we can with equal naturalness fix $\ell_c$ to be comparable with either the Schwarzschild radius of the star or the galaxy extension. 
Indeed, these two are the characteristic scales of the system. On the other hand if we were dealing with a single star in an empty Universe, it would be natural to select $\ell_c$ proportional to the Schwarzschild radius of the star. 
Therefore, conceptually there is nothing wrong in selecting the free scale to be proportional to the galaxy extension, and actually it seems the natural choice whether we are interested to the global properties of the galaxies.  
Furthermore, in conformal gravity we have an extra scalar field, the dilaton, that does not propagate (the perturbation can always be fixed to zero by the mean of conformal symmetry), but satisfies its on equation of motion whose solutions show up extra scales simply because of dimensional reasons and in accordance with the Mach's mechanical view of the Universe. In other words, the dilaton is responsible for the gravitational interaction from small to large distances through the presence of pole-like singularities, which are weighted by dimensional parameters, in the solution of its equation of motion.

The arguments above have an observational counterpart in the Tully-Fisher relation that relates the asymptotic velocity of a probe star to the Newton's constant, the mass of the galaxy, and the Milgrom's parameter $a_0$, namely 
\be
v^4 =  a_0 G M \, , \quad [ a_0 ] = L \, T^{-2} \, , \quad [ G ] = L^3 \, M^{-1} \,T^{-2} \, ,
\label{v4}
\ee
where $M = N^*M_{\odot} + M_{\rm HI}$, $M_{\rm HI}$ is the mass of the Helium gas (see next section for more details). 
Comparing the letter expression (\ref{v4}) with (\ref{vinfinity}) we finally get:
\be
\gamma^* = \sqrt{ \frac{4 a_0}{G M }} \, , \quad [ \gamma^*] = L^{-1} \, , 
\label{gammavalue}
\ee
which depends on the mass of the galaxy whether we assume $a_0$ to be a universal constant. 

For the value of $a_0$ obtained by fitting the galactic rotation curves with the MOND theory \cite{Milgrom:1983zz}, namely $a_0 = 1.2 \times 10^{-10} {\rm m} \, {\rm s}^{-2}$, and for a galaxy made of $10^{12}$ solar mass stars we get:
\be
 \gamma^* \approx 10^{-21} {\rm m}^{-1} \, \quad \Longrightarrow \quad \ell_c \approx 10^{21} {\rm m} \, .
\ee

In conformal gravity $\gamma^*$ is one of the two free parameters to be obtained by fitting the observational data and assuming dependence on the mass of the entire galaxy like in (\ref{gammavalue}). 

In the next section we will get a universal vale for $a_0$ from our model fitting $175$ galaxies.

\section{ Fitting of the galactic rotation curves and universality}

In order to completely specify the velocity square (\ref{vtot}), 
we need: $N^*$ (the number of stars in the galaxy), $R_0$ (the effective scale of the galactic disk), and the free scale in our model, namely $\gamma^*$.
Moreover, 
we have to consider the contribution to the velocity due to the gas Helium (HI). If we apply to the HI the disk model with exponential profile, 
the contribution of HI to $v^2$ will be described by the same formula (\ref{vtot}). Therefore, the total $v^2$ reads:
\be
v^2_{\rm tot}=v^2(N^*, R_0, \gamma^*) + v^2(N_{\rm HI}, R_{\rm H0}, \gamma^*),
\ee
where $N_{\rm HI}= M_{\rm HI}/M_{\odot}$ represents the fraction of total mass of the HI gas respect to the solar mass and $R_{\rm H0}$ is the effective radius of the HI gas' cloud.

%
%

In our analysis we used the data from the SPARC database \cite{Lelli:2016zqa} that includes: the rotation-curves data, which the reader can find in the plots in Appendix \ref{AC}, the total luminosity ratio $L/L_{\odot}$, and the disk radius $R_0({\rm kpc})$ for 175 galaxies (see Appendix \ref{AB}). The database includes also $M_{\rm HI}$, while $R_{\rm H0}$ will be determined shortly. 
Of course, the mass $M_{\odot}$ and the luminosity $L_{\odot}$ of the sun, and the luminosity of all the galaxies $L$ are known observed quantities. All these parameters are given in Appendix \ref{AB}.

The number of stars $N^*$ is related to the mass to luminosity ratio $M/L$, which is our second fitting parameter, the ration $M_{\odot}/L_{\odot}$, and the ration $L/L_{\odot}$, namely 
\be
N^* = \frac{M}{M_\odot} = \frac{\frac{M}{L}}{\frac{M_\odot}{L_\odot}} \, \frac{L}{L_\odot} \, , 
\ee
in which $M_\odot/L_\odot$, and $L/L_\odot$ are known and given in the table in Appendix \ref{AB}.
Therefore, fitting $M/L$ is equivalent to the fitting of $N^*$. 
Since we assume that there is no dark matter, the fitting results of $M/L$ should be close to $1$ rather than over $10$ like in Newtonian dynamics.

In the database \cite{Lelli:2016zqa} we can also find the mass $M_{\rm HI}$. However, in order to also include the amount of primordial Helium, we have to multiply HI times the factor $1.4$. Therefore, the total amount of Helium is:
\be
M^{\rm TOT}_{\rm HI} = 1.4 \, M_{\rm HI} \, .
\ee

In the SPARC database \cite{Lelli:2016zqa} one can find the the radius $R_{\rm H}$ defined to be one for which the density of HI is equal to the value $M_{\odot}/{\rm pc}^2$.
Therefore, we can infer about the effective radius $R_{\rm H0}$ of the Helium gas using the exponential density profile (\ref{rho}) and (\ref{sigma0}), 
\be\label{rh0}
\Sigma_{\rm H0}e^{-R_{\rm H}/R_{\rm H0}}=\frac{N_{\rm HI}}{2\pi R^2_{\rm H0}}e^{-R_{\rm H}/R_{\rm H0}}=\frac{1}{{\rm pc}^2}
\quad \Longrightarrow \quad R_{\rm H0} \, , 
\ee
where the parameters $N_{\rm HI}$, which can be identified with the dimensionless quantity $M_{\rm HI}$, is available in the Appendix \ref{AB}. 
However, equation (\ref{rh0}) is ambiguous because it usually has two solutions. Moreover, for some galaxies, equation (\ref{rh0}) has no solutions, which implies that for these galaxies the measurements of $N_{\rm HI}$ and $R_{\rm H}$ are not accurate enough or the distribution of HI does not fit the disk model properly. Therefore, we choose $R_{\rm H0}=4 R_0$ as an effective radius of the HI disk consistently with other papers in literature \cite{Mannheim:2010xw,Li:2019ksm}.

The results for the fitting parameters $M/L$ and $\gamma^*$ are given in the Appendix \ref{AB}, 
while the fitting of the rotation curves are displayed in the Appendix \ref{AC}. 

The fitting results show that our model fits the rotation velocity data for most of the typical spiral galaxies (including S0, Sa, Sb, Sc, Sab, Sbc, and Scd type) and it fits very well some late spiral type galaxies (Sd, Sdm, and Sm). In particular for the velocity data at large scale ($R>2R_0$).

 As we expected, the fitting results for the mass to luminosity ratio (of luminous mass) are close to $1$. Moreover, in the plots in Appendix \ref{AC}, we can see that the Newtonian contribution dominates the rotation velocity at small scale ($R\lesssim 2R_0$), while the conformally modified geometry determines the the value of the velocity square asymptotically.  Our model (\ref{vtot}) interpolates between the two regimes.

However, there are some galaxies to which our model cannot fit very well. 

This is the case of the galaxies NGC3949, NGC3953, and NGC4051. However, for such galaxies we have only few data and in particular we lack of data points at large radius. In this case the fitting results for $\gamma^*$ is actually 0.

For some spiral galaxies, e.g. NGC2955, NGC5005, NGC6195, UGC2916, UGC3546, UGC5253, and UGC11914,  the rotation velocity data tend to be flat at very small scale ($R\ll 2R_0$). Therefore, we think that the rotation curves cannot be consistent with the exponential profile for the matter density adopted.

For the irregular galaxies, Im (irregular Magellanic), BCD (irregular blue compact dwarf), and weak spiral types (Sm, Sd, and, Sdm), for instance: CamB, DDO161, F574-2, NGC2366, NGC3741, NGC4068, PGC51017, UGC2455, UGC4483, some fits are bad and usually the fitting results of the mass to luminosity ratio are anomalously small. However, this should be related to the irregular mass distribution of these galaxies that affects the irregular motion of matters.

\begin{figure}[ht]
	\centering
	\includegraphics[width=0.5\linewidth]{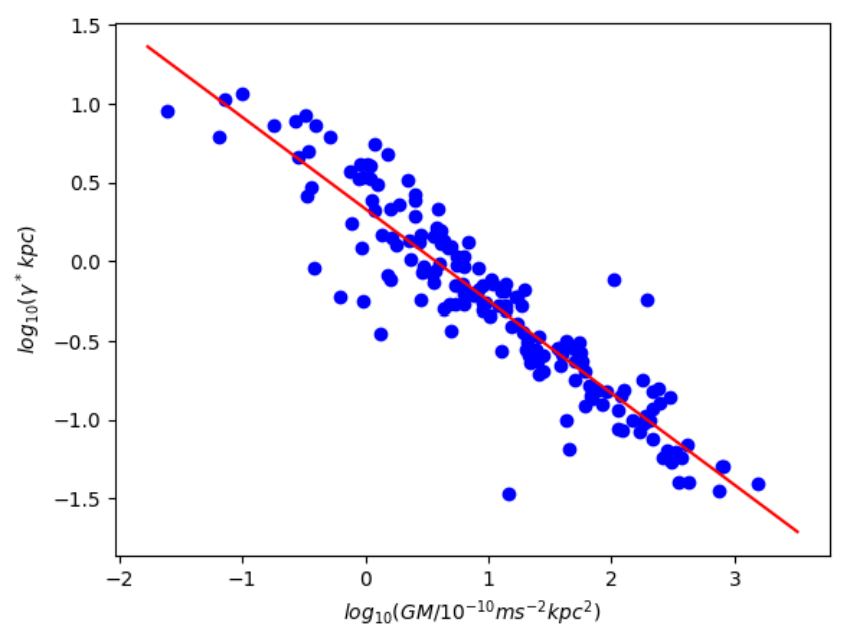}
	\caption{This plot shows the fitting of the relation between $\gamma^*$ and $M$. The fitting function is $y=k(b-x)$, where $y=\log((\gamma^*)\cdot {\rm kpc})$, $x=\log(GM/10^{-10}{\rm ms}^{-2}{\rm kpc}^2)$ and $b=\log(4 a_0/10^{-10}{\rm m s}^{-2})$. The results for the two fitting parameters are: $k=0.582$ and $b=0.573$.}
	\label{tf}
\end{figure}

Finally, having at our disposal the fittings values for $\gamma^*$ and $M/L$ ($L$ is an obseved quantity) we can now extract the {\em universal} parameter $a_0$ using the Tully-Fisher relation (\ref{gammavalue}). 
The total mass in (\ref{gammavalue}) consists of the two contributions, stars and Helium, namely 
\be
M=L\cdot \left(\frac{ M }{L} \right) +1.4 \, M_{\rm HI} \, .
\ee
Let us consider the following generalization of equation (\ref{gammavalue}), namely 
\be
\gamma^* = \left( \frac{4 a_0}{G M } \right)^k \, , 
\label{gammavalue2}
\ee
where the constant $k$ has to be determined by means of the fitting. Hence, taking the ``log'' of both sides we get:
\be
\log \gamma^* = k \left( \log 4 a_0 -  \log GM \right) \, ,
\ee
in which the fitting parameters are $a_0$ and $k$. 
The fitting results are shown in Fig.\ref{tf} (notice that we removed the seven points for which $\gamma^*=0$), 
where 
\be
\gamma^*\propto M^{-0.582} \, ,  \quad  \boxed{a_0= 0.935 \times10^{-10} {\rm m/s}^2= 9.35 \times10^{-11} {\rm m/s}^2 } \, . 
\ee
The $3\sigma$ confidence intervals of $k$ and $a_0$ are: $0.582 \pm 0.057$ and $(9.35 \pm 2.22)\times10^{-11}{\rm m/s}^2$ respectively. Notice that according to (\ref{gammavalue}) $k$ is compatible with $1/ 2$.


\section*{Conclusions}
We provided a geometrical mechanism capable to overcome the long standing issue of the galactic rotation curves without need of any kind of exotic dark matter. We are aware that dark matter is a proposal to rid out of multiple issues in cosmology and astrophysics while there is no need if it in the colliders' physics, but we found extremely interesting the outcome of this project from both the theoretical and observational sides. From the theoretical point of view the simple scalar-tensor Einstein's theory of gravity provides a kind of non-modified gravitational theory ghost-free and free of other instabilities. Indeed, the presence of the dilaton field on one side allows for other vacua without introducing other propagating degrees of freedom, on the other side introduces unattainable spacetime singularities that drastically modify the asymptotic spacetime structure from the micro to the macro.

Specifically, the effective Newtonian gravitational force, to which the stars of the galaxy are subject, is obtained starting from a ``unique'' (the metric depends only on one extra scale $\ell_{\rm c} = 2/\gamma^*$, see section (\ref{Unique})) spacetime geometry (\ref{Qmetric}) or (\ref{Qmetricr}) (in two different coordinate systems) for a single star and summing over all the stars in the galaxy. 
The effective potential have the expected asymptotic logarithmic behaviour characteristic of the minimal confinement, and the velocity turns out to be constant (see formulas (\ref{v2r}) or (\ref{v2rlr}) and (\ref{vinfinity})) at large distance from the galactic center in agreement with the Tully-Fisher relation.

In force of the effective gravitational potential with logarithmic asymptotic behaviour we derived for a single source, we integrated on all the stars of the galaxy with exponential density profile to end up with the total potential. Hence, we obtained the orbital velocity of a probe star in the gravitational field of all the other stars in the galaxy (see (\ref{vtot})). 
Afterwards, we tested the theory with $175$ galaxies making a fit of the parameters: $\gamma^*$ and the mass over the luminosity ratio. The outcome of the fits is given in the appendix (\ref{AC}).
One can notice that the fitting results for the ratio $M/L$ turned out to be close to $1$ consistently with the absence of dark matter.

Finally, using the observational Tully-Fisher relation we got the value for the universal parameter $a_0 = (9.35 \pm 2.22) \times 10^{-11} \, {\rm m/s^2}$. 


\acknowledgments
This work was supported by the Basic Research Program of the Science, Technology, and Innovation Commission of Shenzhen Municipality (grant no. JCYJ20180302174206969). 

\appendix 

\section{Radial geodesic equations in the metric (\ref{Qmetricr})} \label{RadialGeor}
We here derive the radial geodesic equation for massless and conformally coupled particles in the metric (\ref{Qmetricr}).

\subsection{Massless particles} \label{RadialGeor0}
In this section we derive the radial geodesic equation for light in the metric (\ref{Qmetricr}). 
Since like (\ref{Qmetric}) also (\ref{Qmetricr}) is independent on the $t$- and $\varphi$- coordinates, 
according to (\ref{CL}) the following quantities are conserved,  
\be
&& e = - \xi \cdot u 
= - \hat{g}_{ t t } u^t = 
Q^2(r) \left(  1 - \frac{2 M Q^2(r)}{r} \right) \frac{d t}{d \lambda} = Q^2(r) \left(  1 - \frac{2 M Q^2(r)}{r} \right)  \dot{t} \, , 
\label{Ldottr} \\ 
&& \ell = \eta \cdot u = \eta^\alpha u^\beta \hat{g}_{\alpha \beta} =  \hat{g}_{ \phi \beta} u^\beta =  \hat{g}_{ \phi \phi } u^\phi = 
r^2 \sin^2 \theta \, \dot{\varphi} \, ,
\label{Ldotphir}
\ee 
where we introduce the null vector 
\be
u^{\alpha} = \frac{d x^\alpha}{d \lambda} 
\ee
that satisfies 
\be
u \cdot u = \hat{g}_{\alpha \beta}  \frac{d x^\alpha}{d \lambda}  \frac{d x^\beta}{d \lambda} = 0 \, .
\label{NullVr}
\ee
From~(\ref{NullVr}) in the equatorial plane (i.e. $\theta = \pi/2$), we get the following equation
\be
- Q^2(r) \left(  1 - \frac{2 M Q^2(r)}{r} \right) \dot{t}^2 + \frac{\dot{r}^2}{Q^2(r) \left(  1 - \frac{2 M Q^2(r)}{r} \right)} 
+ r^2 \sin^2 \theta \dot{\varphi}^2 = 0 \, .
\label{uuA}
\ee
Solving (\ref{Ldottr}) for $\dot{t}$ and (\ref{Ldotphir}) for $\dot{\varphi}$ and replacing the results in (\ref{uuA}), the radial geodesic equation ($\ell =0$) reads:
\be
-  \frac{e^2}{Q(r)^2 \left(1 - \frac{2M Q(r)}{r} \right)} 
+ \frac{\dot{r}^2}{Q^2(r) \left(  1 - \frac{2 M Q^2(r)}{r} \right)} 
= 0 \quad \Longrightarrow \quad -  e^2 
+\dot{r}^2 =0 \, \quad \Longrightarrow \quad | \dot{r} | = e 
 \, , 
\label{uu2r}
\ee
which coincides with (\ref{Geom2}). 
 
 \subsection{Conformally coupled massive particles} \label{RadialGeor2}
We here study the radial geodesic equations for conformally coupled particles in the metric (\ref{Qmetricr}), namely for the metric in the radial coordinate $r$. 
The Lagrangian for a conformally coupled particle reads: 
\be
L_{\rm cp} = - \sqrt{ - f^2 \phi^2 \hat{g}_{\mu\nu} \dot{x}^\mu \dot{x}^\nu } \, , 
\label{Lcp3}
\ee
and the translation invariance in the time-like coordinate $t$ implies: 
\be
\frac{\partial L_{\rm cp}}{\partial \dot{t} } = - \frac{f^2 \phi^2 \hat{g}_{tt} \dot{t}}{L_{\rm cp}} = {\rm const.} = - E \quad \Longrightarrow \quad \dot{t} = \frac{L_{\rm cp} E }{f^2 \phi^2 \hat{g}_{tt}} . 
\label{ConstE2}
\ee
In the proper time gauge,
\be
\frac{d \hat{s}^2}{d \tau^2} = - 1 
\quad \Longrightarrow \quad 
\hat{g}_{\mu\nu} \dot{x}^\mu \dot{x}^\nu = - 1 \quad {\rm and} 
\quad 
L_{\rm cp}= - f \phi 
\quad \Longrightarrow \quad  \dot{t} = - \frac{ E }{f \phi \, \hat{g}_{tt}}  \, .
\label{PTG2}
\ee
Therefore, 
\be
&&\hat{g}_{tt} \dot{t}^2 + \hat{g}_{rr} \dot{r}^2 = -1 \quad \Longrightarrow \quad 
\hat{g}_{tt} \left( - \frac{  E}{f \, \phi \, \hat{g}_{tt}} \right)^2 + \hat{g}_{rr} \dot{r}^2 = -1
\quad \Longrightarrow \quad 
 \frac{E^2}{f^2 \phi^2 } + \hat{g}_{tt} \hat{g}_{rr} \dot{r}^2 = - \hat{g}_{tt} \nonumber \\
 &&  \Longrightarrow \quad \frac{E^2}{f^2 \kappa_4^{-1} Q(r)^{-2} } - \dot{r}^2 =  Q(r)^2 \left( 1 - \frac{2 G M}{r} Q(r) \right) 
 \Longrightarrow \quad 
 e^2 \, Q(r)^2 - \dot{r}^2 =  Q(r)^2 \left( 1 - \frac{2 G M}{r} Q(r) \right) \nonumber \\
 && \Longrightarrow \quad 
  \dot{r}^2 =  Q(r)^2 \left( e^2 - 1 + \frac{2 G M}{r} Q(r) \right) \, , 
\ee
which coincides with (\ref{confrar}).

\clearpage
\section{Galactic parameters for $175$ galaxies}\label{AB}
In this section we remind the main data from the SPARC database \cite{Lelli:2016zqa} need for the fits of the square velocity (\ref{vtot}), and we list the values of $\gamma^*$ and $M/L$ that turn out from our fits for $175$ galaxies. 
\vspace{-5pt}
\begin{table}[H]
	\centering
	\label{Tab01}
	\begin{ruledtabular}
		\begin{tabular}{cccccccc}	
			\toprule
			
		Galaxy & Hubble &	Distance & $L$ &	$R_0$ &	$M_{\rm HI}$ & $(M/L)_{\rm stars}$ & $\gamma^*$  \\
			
			
        Name&Type &(Mpc)&($10^{9}L_{\odot}$) &(kpc)&$(10^{9}M_{\odot})$ & $(M_{\odot}/L_{\odot})$ & $({\rm kpc}^{-1})$ \\ \hline
        CamB & Im & 3.36 & 0.075 & 0.47 & 0.012 & 0.0883 & 8.94 \\
        D512-2 & Im & 15.2 & 0.325 & 1.24 & 0.081 & 1.75 & 0.556 \\
        D564-8 & Im & 8.79 & 0.033 & 0.61 & 0.029 & 0.295 & 10.6 \\
        D631-7 & Im & 7.72 & 0.196 & 0.7 & 0.29 & 1.19 & 4.13 \\ 
        DDO064 & Im & 6.8 & 0.157 & 0.69 & 0.211 & 0.411 & 6.18 \\ 
        DDO154 & Im & 4.04 & 0.053 & 0.37 & 0.275 & 1.09 & 0.6 \\
        DDO161 & Im & 7.5 & 0.548 & 1.22 & 1.378 & 0.0892 & 0.571 \\
        DDO168 & Im & 4.25 & 0.191 & 1.02 & 0.413 & 1.01 & 3.32 \\
        DDO170 & Im & 15.4 & 0.543 & 1.95 & 0.735 & 1.03 & 1.36 \\
        ESO079-G014 & Sbc & 28.7 & 51.733 & 5.08 & 3.14 & 0.657 & 0.311 \\
        ESO116-G012 & Sd & 13 & 4.292 & 1.51 & 1.083 & 0.676 & 1.08 \\
        ESO444-G084 & Im & 4.83 & 0.071 & 0.46 & 0.135 & 1.24 & 7.35 \\
        ESO563-G021 & Sbc & 60.8 & 311.177 & 5.45 & 24.298 & 0.56 & 0.139 \\
        F561-1 & Sm & 66.4 & 4.077 & 2.79 & 1.622 & 0.2 & 0.5 \\
        F563-1 & Sm & 48.9 & 1.903 & 3.52 & 3.2 & 2.71 & 0.725 \\
        F563-V1 & Im & 54 & 1.54 & 3.79 & 0.61 & 1.17 & 0 \\
        F563-V2 & Im & 59.7 & 2.986 & 2.43 & 2.169 & 3.07 & 0.409 \\
        F565-V2 & Im & 51.8 & 0.559 & 2.17 & 0.699 & 1.06 & 3.28 \\
        F567-2 & Sm & 79 & 2.134 & 3.08 & 2.449 & 0.5 & 0.68 \\
        F568-1 & Sc & 90.7 & 6.252 & 5.18 & 4.498 & 4.49 & 0.283 \\
        F568-3 & Sd & 82.4 & 8.346 & 4.99 & 3.195 & 1.81 & 0.257 \\
        F568-V1 & Sd & 80.6 & 3.825 & 2.85 & 2.491 & 2.64 & 0.362 \\
        F571-8 & Sc & 53.3 & 10.164 & 3.56 & 1.782 & 1.099 & 0.662 \\
        F571-V1 & Sd & 80.1 & 1.849 & 2.47 & 1.217 & 0.633 & 1.57 \\
        F574-1 & Sd & 96.8 & 6.537 & 4.46 & 3.524 & 1.9 & 0.277 \\
        F574-2 & Sm & 89.1 & 2.877 & 3.76 & 1.701 & 0.0654 & 0.871 \\
        F579-V1 & Sc & 89.5 & 11.848 & 3.37 & 2.245 & 1.4 & 0.205 \\
        F583-1 & Sm & 35.4 & 0.986 & 2.36 & 2.126 & 1.51 & 0.922 \\
        F583-4 & Sc & 53.3 & 1.715 & 1.93 & 0.641 & 0.955 & 0.736 \\
        IC2574 & Sm & 3.91 & 1.016 & 2.78 & 1.036 & 0.319 & 2.43 \\
        IC4202 & Sbc & 100.4 & 179.749 & 4.78 & 12.326 & 0.64 & 0.107 \\
        KK98-251 & Im & 6.8 & 0.085 & 1.34 & 0.115 & 0.789 & 8.4 \\
        NGC0024 & Sc & 7.3 & 3.889 & 1.34 & 0.676 & 1.49 & 0.542 \\
        NGC0055 & Sm & 2.11 & 4.628 & 6.11 & 1.565 & 2.67 & 0.329 \\ 
        NGC0100 & Scd & 13.5 & 3.232 & 1.66 & 1.99 & 0.345 & 1.07 \\
        NGC0247 & Sd & 3.7 & 7.332 & 3.74 & 1.746 & 1 & 0.489 \\
        NGC0289 & Sbc & 20.8 & 72.065 & 6.74 & 27.469 & 1.122 & 0.0833 \\
        NGC0300 & Sd & 2.08 & 2.922 & 1.75 & 0.936 & 0.612 & 1.35 \\
        NGC0801 & Sc & 80.7 & 312.57 & 8.72 & 23.201 & 0.847 & 0.04 \\
        NGC0891 & Sb & 9.91 & 138.34 & 2.55 & 4.462 & 0.532 & 0.0876 \\
        NGC1003 & Scd & 11.4 & 6.82 & 1.61 & 5.88 & 0.121 & 0.271 \\
        NGC1090 & Sbc & 37 & 72.045 & 3.53 & 8.783 & 0.513 & 0.147 \\
        NGC1705 & BCD & 5.73 & 0.533 & 0.39 & 0.139 & 1.22 & 2.09 \\
        NGC2366 & Im & 3.27 & 0.236 & 0.65 & 0.647 & 0.15 & 0.35 \\
        NGC2403 & Scd & 3.16 & 10.041 & 1.39 & 3.199 & 0.446 & 0.646 \\
        \bottomrule
	\end{tabular}
	\end{ruledtabular}
\end{table}  
        \clearpage
\begin{table}[H]
	\centering
	\label{Tab02}
	\begin{ruledtabular}
		\begin{tabular}{cccccccc}	
			\toprule
			
		Galaxy & Hubble &	Distance & $L$ &	$R_0$ &	$M_{\rm HI}$ & $(M/L)_{\rm stars}$ & $\gamma^*$  \\
			
			
        Name&Type &(Mpc)&($10^{9}L_{\odot}$) &(kpc)&$(10^{9}M_{\odot})$ & $(M_{\odot}/L_{\odot})$ & $({\rm kpc}^{-1})$ \\ \hline        
        NGC2683 & Sb & 9.81 & 80.415 & 2.18 & 1.406 & 0.559 & 0.165 \\
        NGC2841 & Sb & 14.1 & 188.121 & 3.64 & 9.775 & 0.822 & 0.158 \\
        NGC2903 & Sbc & 6.6 & 81.863 & 2.33 & 2.552 & 0.684 & 0.126 \\
        NGC2915 & BCD & 4.06 & 0.641 & 0.55 & 0.508 & 0.286 & 3.05 \\
         NGC2955 & Sb & 97.9 & 319.422 & 18.76 & 28.949 & 3.3 & 0.0396 \\
        NGC2976 & Sc & 3.58 & 3.371 & 1.01 & 0.172 & 0.498 & 1.33 \\
        NGC2998 & Sc & 68.1 & 150.902 & 6.2 & 23.451 & 1 & 0.0576 \\  	
        NGC3109 & Sm & 1.33 & 0.194 & 1.56 & 0.477 & 0.854 & 5.56 \\
        NGC3198 & Sc & 13.8 & 38.279 & 3.14 & 10.869 & 0.533 & 0.179 \\
        NGC3521 & Sbc & 7.7 & 84.836 & 2.4 & 4.154 & 0.808 & 0.768 \\
        NGC3726 & Sc & 18 & 70.234 & 3.4 & 6.473 & 0.315 & 0.292 \\
        NGC3741 & Im & 3.21 & 0.028 & 0.2 & 0.182 & 0.638 & 0.915 \\
        NGC3769 & Sb & 18 & 18.679 & 3.38 & 5.529 & 1.2 & 0.0986 \\
        NGC3877 & Sc & 18 & 72.535 & 2.53 & 1.483 & 0.33 & 0.288 \\
        NGC3893 & Sc & 18 & 58.525 & 2.38 & 5.799 & 0.668 & 0.143 \\
        NGC3917 & Scd & 18 & 21.966 & 2.63 & 1.888 & 0.47 & 0.523 \\
        NGC3949 & Sbc & 18 & 38.067 & 3.59 & 3.371 & 1.74 & 0 \\
        NGC3953 & Sbc & 18 & 141.301 & 4.89 & 2.832 & 1.08 & 0 \\
        NGC3972 & Sbc & 18 & 14.353 & 2.18 & 1.214 & 0.545 & 0.647 \\
        NGC3992 & Sbc & 23.7 & 226.932 & 4.96 & 16.599 & 0.574 & 0.117 \\
        NGC4010 & Sd & 18 & 17.193 & 2.81 & 2.832 & 0.465 & 0.591 \\
        NGC4013 & Sb & 18 & 79.094 & 3.53 & 2.967 & 0.632 & 0.154 \\
        NGC4051 & Sbc & 18 & 95.268 & 4.65 & 2.697 & 7.62 & 0 \\
        NGC4068 & Im & 4.37 & 0.236 & 0.59 & 0.154 & 0.118 & 5.04 \\
        NGC4085 & Sc & 18 & 21.724 & 1.65 & 1.349 & 0.367 & 0.526 \\
        NGC4088 & Sbc & 18 & 107.286 & 2.58 & 8.226 & 0.227 & 0.236 \\
        NGC4100 & Sbc & 18 & 59.394 & 2.15 & 3.102 & 0.431 & 0.318 \\
        NGC4138 & S0 & 18 & 44.111 & 1.51 & 1.483 & 0.574 & 0.223 \\
        NGC4157 & Sb & 18 & 105.62 & 2.32 & 8.226 & 0.277 & 0.238 \\
        NGC4183 & Scd & 18 & 10.838 & 2.79 & 3.506 & 0.981 & 0.231 \\
        NGC4214 & Im & 2.87 & 1.141 & 0.51 & 0.486 & 0.844 & 1.02 \\
        NGC4217 & Sb & 18 & 85.299 & 2.94 & 2.562 & 0.461 & 0.205 \\
        NGC4389 & Sbc & 18 & 21.328 & 2.79 & 0.539 & 0.367 & 0.53 \\
        NGC4559 & Scd & 9 & 19.377 & 2.1 & 5.811 & 0.356 & 0.259 \\
        NGC5005 & Sbc & 16.9 & 178.72 & 9.45 & 1.28 & 4.63 & 0 \\
        NGC5033 & Sc & 15.7 & 110.509 & 5.16 & 11.314 & 1.03 & 0.0951 \\
        NGC5055 & Sbc & 9.9 & 152.922 & 3.2 & 11.722 & 0.458 & 0.0863 \\
        NGC5371 & Sbc & 39.7 & 340.393 & 7.44 & 11.18 & 0.593 & 0.0542 \\
        NGC5585 & Sd & 7.06 & 2.943 & 1.53 & 1.683 & 0.51 & 0.939 \\
        NGC5907 & Sc & 17.3 & 175.425 & 5.34 & 21.025 & 0.699 & 0.0755 \\
        NGC5985 & Sb & 39.7 & 208.728 & 7.01 & 11.586 & 1.32 & 0.0692 \\
        NGC6015 & Scd & 17 & 32.129 & 2.3 & 5.834 & 0.609 & 0.261 \\
        NGC6195 & Sb & 127.8 & 391.076 & 13.94 & 20.907 & 1.35 & 0.0506 \\
        NGC6503 & Scd & 6.26 & 12.845 & 2.16 & 1.744 & 0.931 & 0.279 \\
        NGC6674 & Sb & 51.2 & 214.654 & 6.04 & 32.165 & 0.892 & 0.063 \\
        NGC6789 & BCD & 3.52 & 0.1 & 0.31 & 0.017 & 1.67 & 7.75 \\%
        NGC6946 & Scd & 5.52 & 66.173 & 2.44 & 5.67 & 0.533 & 0.123 \\
        NGC7331 & Sb & 14.7 & 250.631 & 5.02 & 11.067 & 0.659 & 0.0632 \\
        NGC7793 & Sd & 3.61 & 7.05 & 1.21 & 0.861 & 0.594 & 0.609 \\
        NGC7814 & Sab & 14.4 & 74.529 & 2.54 & 1.07 & 1.04 & 0.116 \\
        PGC51017 & BCD & 13.6 & 0.155 & 0.53 & 0.201 & 0 & 0 \\
        UGC00128 & Sdm & 64.5 & 12.02 & 5.95 & 7.431 & 1.35 & 0.288 \\
        UGC00191 & Sm & 17.1 & 2.004 & 1.58 & 1.343 & 1.08 & 0.706 \\
        UGC00634 & Sm & 30.9 & 2.989 & 2.45 & 3.663 & 0.748 & 0.774 \\
        \bottomrule
	\end{tabular}
	\end{ruledtabular}
\end{table}      
        \clearpage
\begin{table}[H]
	\centering
	\label{Tab03}
	\begin{ruledtabular}
		\begin{tabular}{cccccccc}	
			\toprule
			
		Galaxy & Hubble &	Distance & $L$ &	$R_0$ &	$M_{\rm HI}$ & $(M/L)_{\rm stars}$ & $\gamma^*$  \\
			
			
        Name&Type &(Mpc)&($10^{9}L_{\odot}$) &(kpc)&$(10^{9}M_{\odot})$ & $(M_{\odot}/L_{\odot})$ & $({\rm kpc}^{-1})$ \\ \hline
        UGC00731 & Im & 12.5 & 0.323 & 2.3 & 1.807 & 3.08 & 1.24 \\
        UGC00891 & Sm & 10.2 & 0.374 & 1.43 & 0.428 & 0.464 & 4.01 \\
        UGC01230 & Sm & 53.7 & 7.62 & 4.34 & 6.43 & 2.965 & 0.0649 \\
        UGC01281 & Sdm & 5.27 & 0.353 & 1.63 & 0.294 & 0.915 & 4.12 \\
        UGC02023 & Im & 10.4 & 1.308 & 1.55 & 0.477 & 0.346 & 2.14 \\
        UGC01281 & Sdm & 5.27 & 0.353 & 1.63 & 0.294 & 0.915 & 4.12 \\
        UGC02023 & Im & 10.4 & 1.308 & 1.55 & 0.477 & 0.346 & 2.14 \\
        UGC02259 & Sdm & 10.5 & 1.725 & 1.62 & 0.494 & 2.19 & 0.538 \\
        UGC02455 & Im & 6.92 & 3.649 & 0.99 & 0.803 & 0.0341 & 1.28 \\
        UGC02487 & S0 & 69.1 & 489.955 & 7.89 & 17.963 & 1.12 & 0.0508 \\
        UGC02885 & Sc & 80.6 & 403.525 & 11.4 & 40.075 & 1.18 & 0.0352 \\
        UGC02916 & Sab & 65.4 & 124.153 & 6.15 & 23.273 & 1.33 & 0.0635 \\
        UGC02953 & Sab & 16.5 & 259.518 & 3.55 & 7.678 & 0.55 & 0.15 \\
        UGC03205 & Sab & 50 & 113.642 & 3.19 & 9.677 & 0.659 & 0.154 \\
        UGC03546 & Sa & 28.7 & 101.336 & 3.79 & 2.675 & 0.6 & 0.152 \\
        UGC03580 & Sa & 20.7 & 13.266 & 2.43 & 4.37 & 0.886 & 0.194 \\
        UGC04278 & Sd & 9.51 & 1.307 & 2.21 & 1.116 & 0.882 & 2.14 \\
        UGC04305 & Im & 3.45 & 0.736 & 1.16 & 0.69 & 0.134 & 0.824 \\
        UGC04325 & Sm & 9.6 & 2.026 & 1.86 & 0.678 & 2.64 & 0.497 \\
        UGC04483 & Im & 3.34 & 0.013 & 0.18 & 0.032 & 0.0444 & 6.2 \\
        UGC04499 & Sdm & 12.5 & 1.552 & 1.73 & 1.1 & 0.847 & 0.971 \\
        UGC05005 & Im & 53.7 & 4.1 & 3.2 & 3.093 & 0.359 & 0.908 \\
        UGC05253 & Sab & 22.9 & 171.582 & 8.07 & 16.396 & 1.3 & 0.04 \\
        UGC05414 & Im & 9.4 & 1.123 & 1.47 & 0.574 & 0.479 & 2.31 \\
        UGC05716 & Sm & 21.3 & 0.588 & 1.14 & 1.094 & 0.923 & 0.922 \\
        UGC05721 & Sd & 6.18 & 0.531 & 0.38 & 0.562 & 0.722 & 1.4 \\
        UGC05750 & Sdm & 58.7 & 3.336 & 3.46 & 1.099 & 0.409 & 1.34 \\
        UGC05764 & Im & 7.47 & 0.085 & 1.17 & 0.163 & 6.58 & 2.42 \\
        UGC05829 & Im & 8.64 & 0.564 & 1.99 & 1.023 & 0.624 & 2.63 \\
        UGC05918 & Im & 7.66 & 0.233 & 1.66 & 0.297 & 2.3 & 1.46 \\
        UGC05986 & Sm & 8.63 & 4.695 & 1.67 & 2.667 & 0.824 & 0.725 \\
        UGC05999 & Im & 47.7 & 3.384 & 3.22 & 2.022 & 0.577 & 1.33 \\
        UGC06399 & Sm & 18 & 2.296 & 2.05 & 0.674 & 0.748 & 1.62 \\
        UGC06446 & Sd & 12 & 0.988 & 1.49 & 1.379 & 1.91 & 0.711 \\
        UGC06614 & Sa & 88.7 & 124.35 & 5.1 & 21.888 & 0.434 & 0.142 \\
        UGC06628 & Sm & 15.1 & 3.739 & 2.82 & 1.5 & 0.37 & 0.366 \\
        UGC06667 & Scd & 18 & 1.397 & 5.15 & 0.809 & 7.52 & 0.598 \\
        UGC06786 & S0 & 29.3 & 73.407 & 3.6 & 5.03 & 1.32 & 0.0985 \\
        UGC06787 & Sab & 21.3 & 98.256 & 5.37 & 5.03 & 2.01 & 0.0713 \\%
        UGC06818 & Sm & 18 & 1.588 & 1.39 & 1.079 & 0.157 & 1.94 \\
        UGC06917 & Sm & 18 & 6.832 & 2.76 & 2.023 & 1.18 & 0.39 \\
        UGC06923 & Im & 18 & 2.89 & 1.44 & 0.809 & 0.48 & 1.44 \\
        UGC06930 & Sd & 18 & 8.932 & 3.94 & 3.237 & 1.49 & 0.237 \\
        UGC06973 & Sab & 18 & 53.87 & 1.07 & 1.753 & 0.295 & 0.334 \\
        UGC06983 & Scd & 18 & 5.298 & 3.21 & 2.967 & 1.97 & 0.307 \\
        UGC07089 & Sdm & 18 & 3.585 & 2.26 & 1.214 & 0.333 & 1.29 \\
        UGC07125 & Sm & 19.8 & 2.712 & 3.38 & 4.629 & 1.42 & 0.0337 \\
        UGC07151 & Scd & 6.87 & 2.284 & 1.25 & 0.616 & 0.691 & 0.843 \\
        UGC07232 & Im & 2.83 & 0.113 & 0.29 & 0.046 & 0.57 & 7.29 \\
        UGC07261 & Sdm & 13.1 & 1.753 & 1.2 & 1.388 & 0.827 & 0.534 \\
        UGC07323 & Sdm & 8 & 4.109 & 2.26 & 0.722 & 0.566 & 1.21 \\
        UGC07399 & Sdm & 8.43 & 1.156 & 1.64 & 0.745 & 4.09 & 0.658 \\
        UGC07524 & Sm & 4.74 & 2.436 & 3.46 & 1.779 & 1.96 & 0.453 \\
        UGC07559 & Im & 4.97 & 0.109 & 0.58 & 0.169 & 0.166 & 2.95 \\
        \bottomrule
	\end{tabular}
	\end{ruledtabular}
\end{table}      
        \clearpage
\begin{table}[H]
	\centering
	\label{Tab04}
	\begin{ruledtabular}
		\begin{tabular}{cccccccc}	
			\toprule
			
		Galaxy & Hubble &	Distance & $L$ &	$R_0$ &	$M_{\rm HI}$ & $(M/L)_{\rm stars}$ & $\gamma^*$  \\
			
			
        Name&Type &(Mpc)&($10^{9}L_{\odot}$) &(kpc)&$(10^{9}M_{\odot})$ & $(M_{\odot}/L_{\odot})$ & $({\rm kpc}^{-1})$ \\ \hline
        UGC07577 & Im & 2.59 & 0.045 & 0.9 & 0.044 & 0.172 & 11.5 \\
        UGC07603 & Sd & 4.7 & 0.376 & 0.53 & 0.258 & 0.438 & 3.75 \\
        UGC07608 & Im & 8.21 & 0.264 & 1.5 & 0.535 & 1.16 & 4.77 \\
        UGC07690 & Im & 8.11 & 0.858 & 0.57 & 0.39 & 0.692 & 0.763 \\
        UGC07866 & Im & 4.57 & 0.124 & 0.61 & 0.118 & 0.56 & 2.6 \\
        UGC08286 & Scd & 6.5 & 1.255 & 1.05 & 0.642 & 0.876 & 1.48 \\
        UGC07866 & Im & 4.57 & 0.124 & 0.61 & 0.118 & 0.56 & 2.6 \\
        UGC08286 & Scd & 6.5 & 1.255 & 1.05 & 0.642 & 0.876 & 1.48 \\
        UGC08490 & Sm & 4.65 & 1.017 & 0.67 & 0.72 & 0.999 & 0.854 \\
        UGC08550 & Sd & 6.7 & 0.289 & 0.45 & 0.288 & 0.47 & 1.74 \\
        UGC08699 & Sab & 39.3 & 50.302 & 3.09 & 3.738 & 1.23 & 0.0997 \\
        UGC08837 & Im & 7.21 & 0.501 & 1.72 & 0.32 & 0.462 & 3.42 \\
        UGC09037 & Scd & 83.6 & 68.614 & 4.28 & 19.078 & 0.335 & 0.137 \\
        UGC09133 & Sab & 57.1 & 282.926 & 6.97 & 33.428 & 0.75 & 0.058 \\
        UGC09992 & Im & 10.7 & 0.336 & 1.04 & 0.318 & 0.643 & 1.22 \\
        UGC10310 & Sm & 15.2 & 1.741 & 1.8 & 1.196 & 1.24 & 0.536 \\
        UGC11455 & Scd & 78.6 & 374.322 & 5.93 & 13.335 & 0.415 & 0.127 \\
        UGC11557 & Sdm & 24.2 & 12.101 & 2.75 & 2.605 & 0.215 & 0.704 \\
        UGC11820 & Sm & 18.1 & 0.97 & 2.08 & 1.977 & 1.58 & 0.718 \\
        UGC11914 & Sab & 16.9 & 150.028 & 2.44 & 0.888 & 0.907 & 0.577 \\
        UGC12506 & Scd & 100.6 & 139.571 & 7.38 & 35.556 & 1.32 & 0.0599 \\
        UGC12632 & Sm & 9.77 & 1.301 & 2.42 & 1.744 & 1.58 & 0.62 \\
        UGC12732 & Sm & 13.2 & 1.667 & 1.98 & 3.66 & 0.704 & 0.548 \\
        UGCA281 & BCD & 5.68 & 0.194 & 1.72 & 0.062 & 13.1 & 0 \\
        UGCA442 & Sm & 4.35 & 0.14 & 1.18 & 0.263 & 1.86 & 3.37 \\
        UGCA444 & Im & 0.98 & 0.012 & 0.83 & 0.067 & 8.9 & 4.57 \\
	\bottomrule
	\end{tabular}
	\end{ruledtabular}
\end{table}

\section{Fitting the galactic rotation curves of $175$ galaxies}\label{AC}
We hereby provide the fits for the galactic orbital velocity  (in km/s) as a function of the physical radial distance (in kpc) for $175$ galaxies. In each plot, the dashed (blue-)curve represents the Newtonian contribution to the velocity square, namely the first term in (\ref{vtot}), while the (yellow-)dot-dashed curve shows only the modification due to the conformal rescaling, namely only the second contribution in (\ref{vtot}).

\vspace{-5pt}
\begin{figure}[H]
	\centering
	\includegraphics[width=\linewidth]{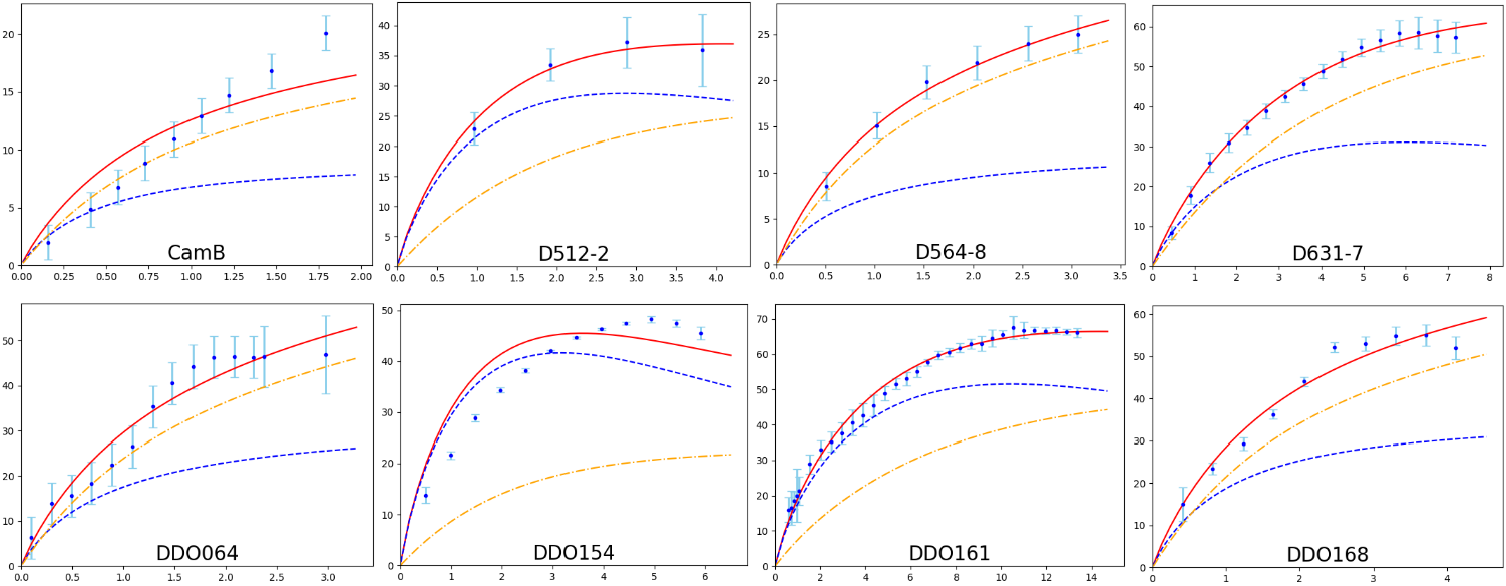}
\end{figure}

\begin{figure}[H]
	\centering
	\includegraphics[width=\linewidth]{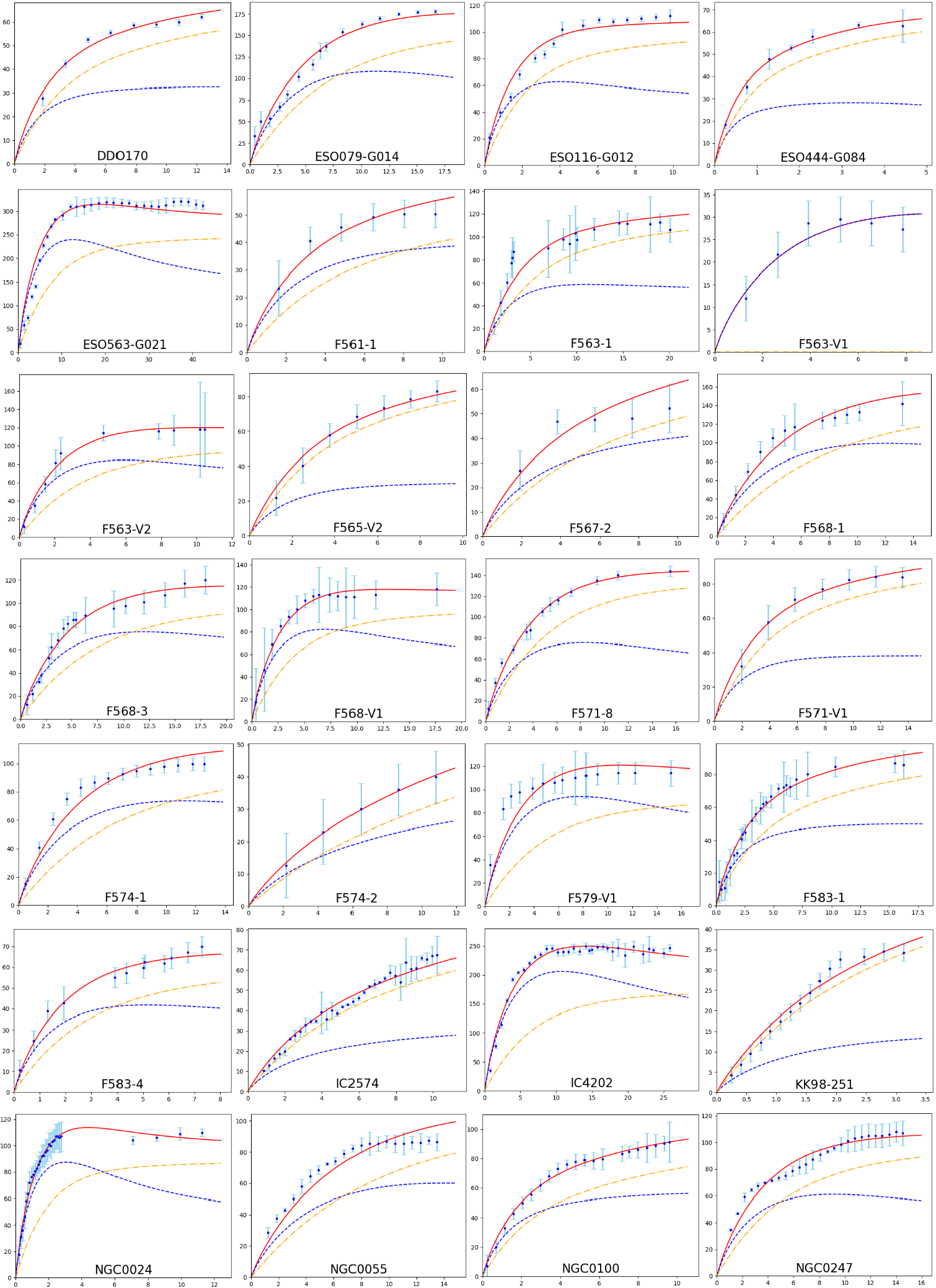}
\end{figure}

\begin{figure}[H]
	\centering
	\includegraphics[width=\linewidth]{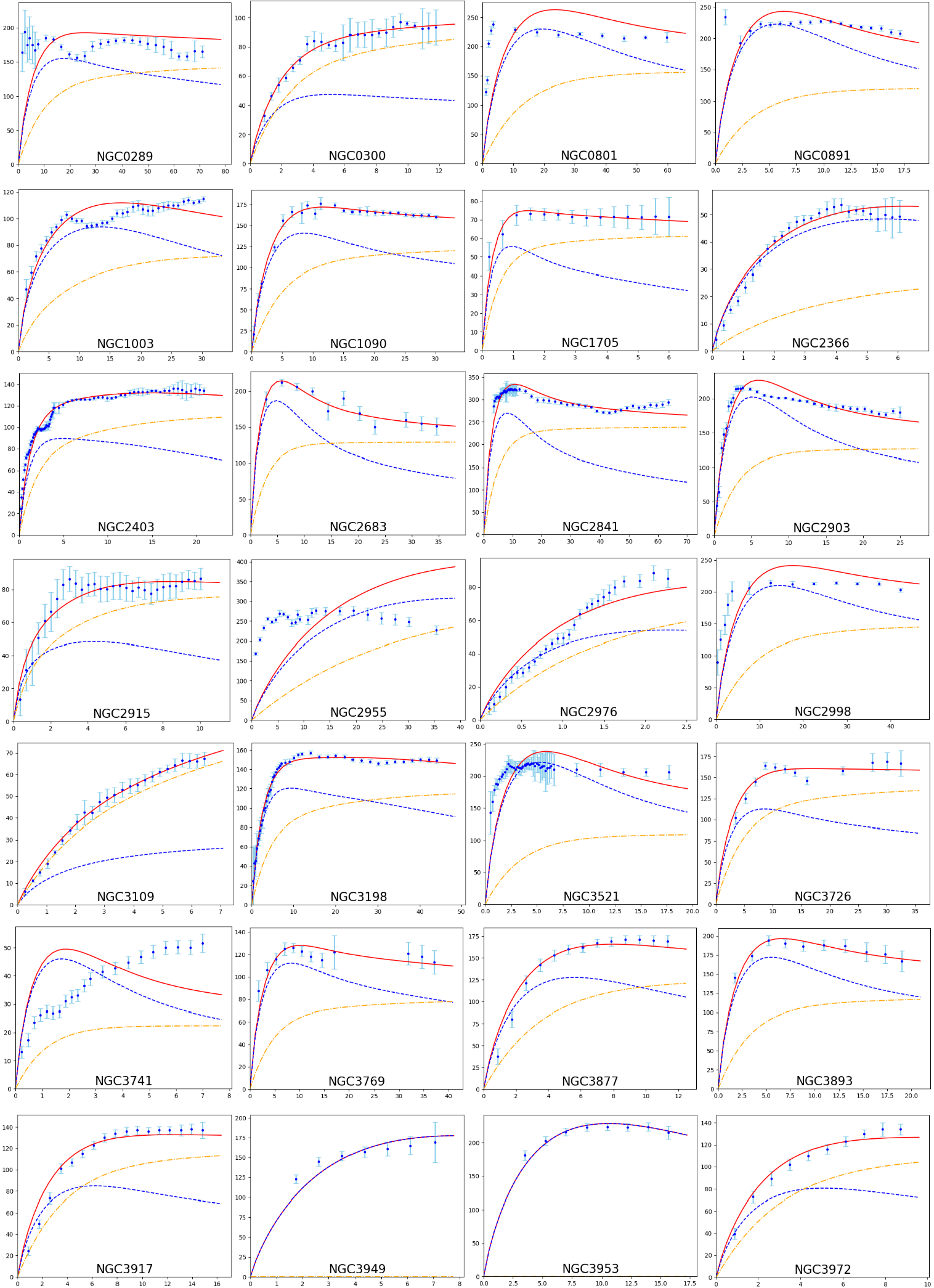}
\end{figure}

\begin{figure}[H]
	\centering
	\includegraphics[width=\linewidth]{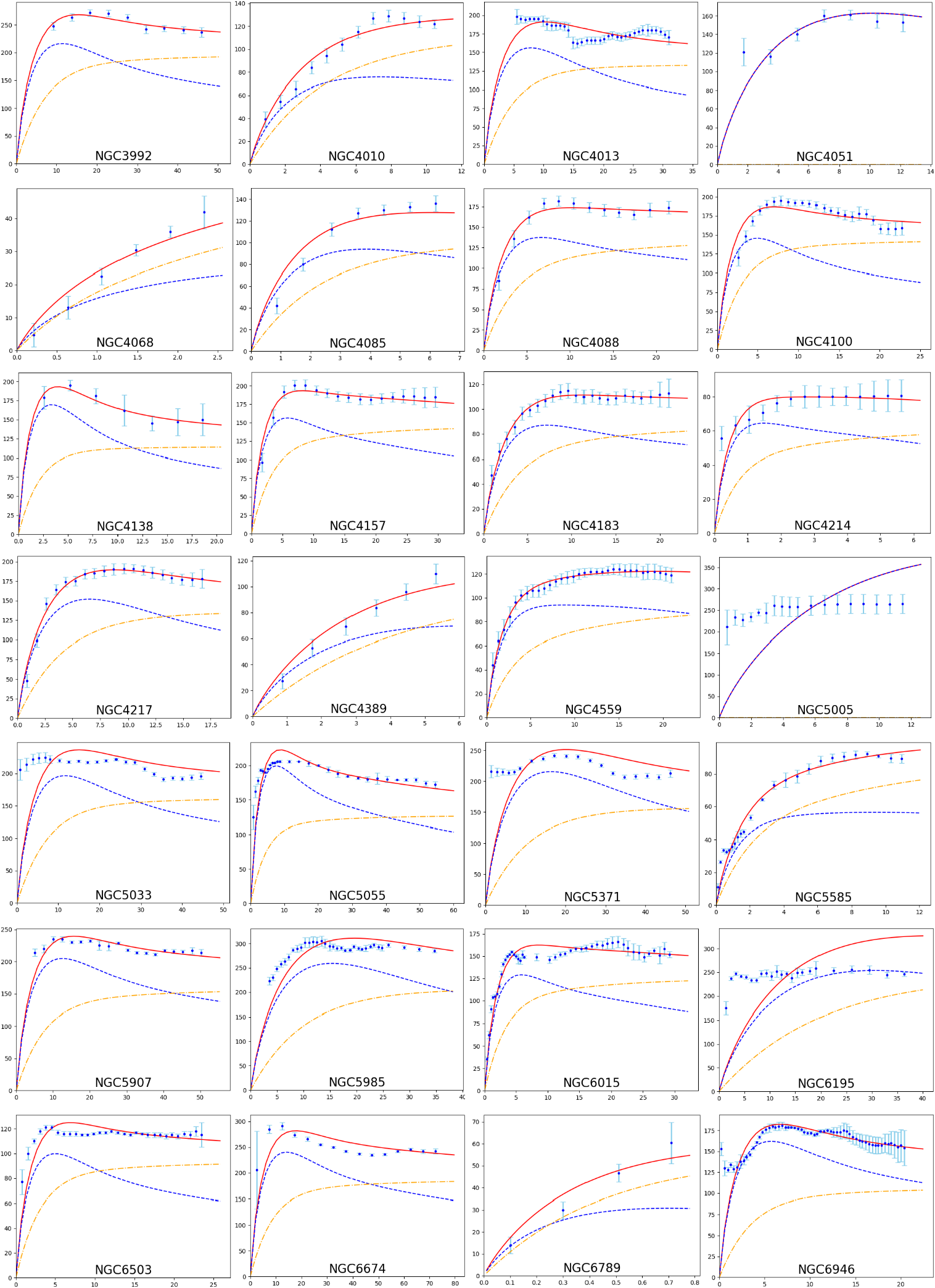}
\end{figure}

\begin{figure}[H]
	\centering
	\includegraphics[width=\linewidth]{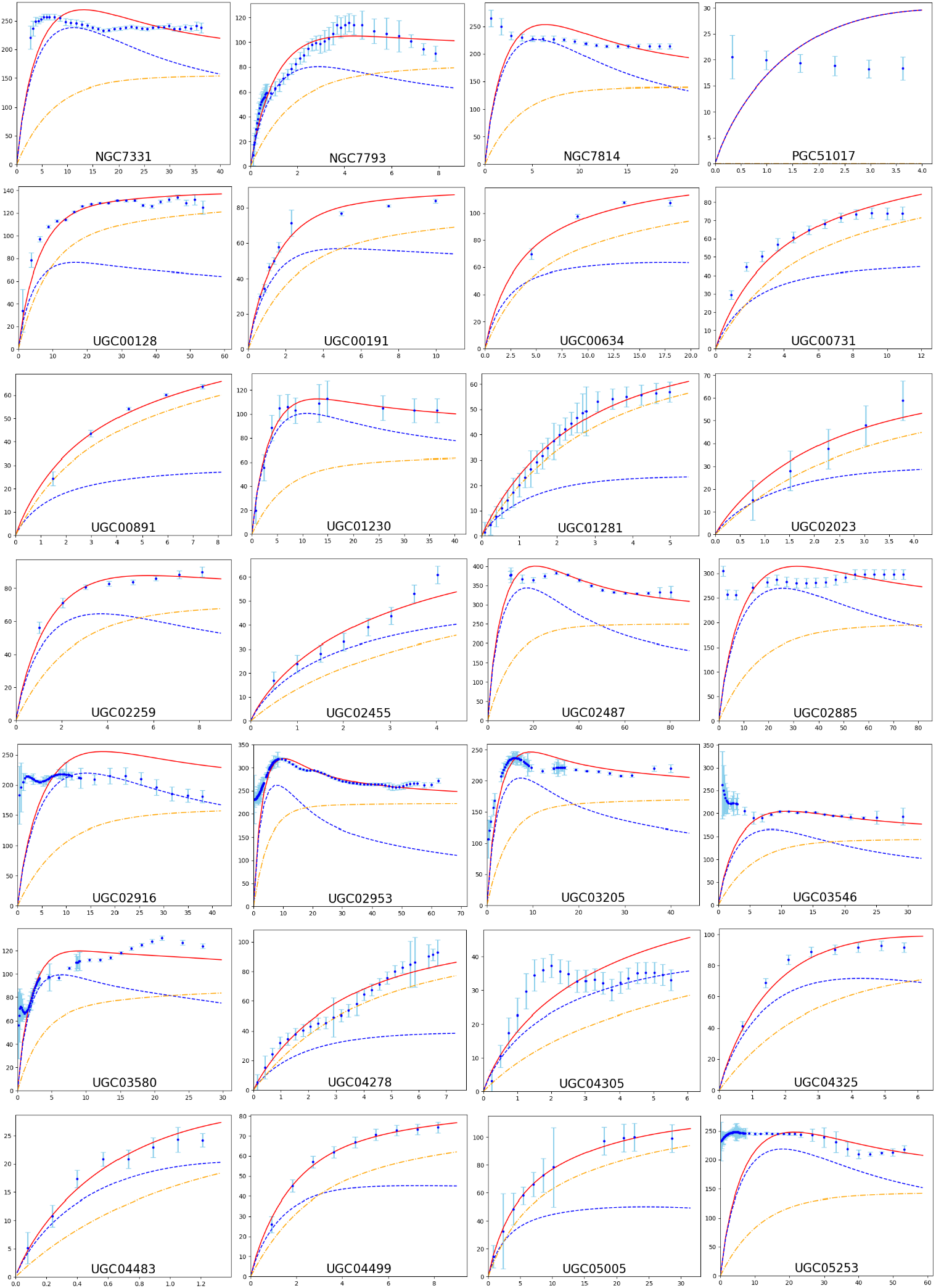}
\end{figure}

\begin{figure}[H]
	\centering
	\includegraphics[width=\linewidth]{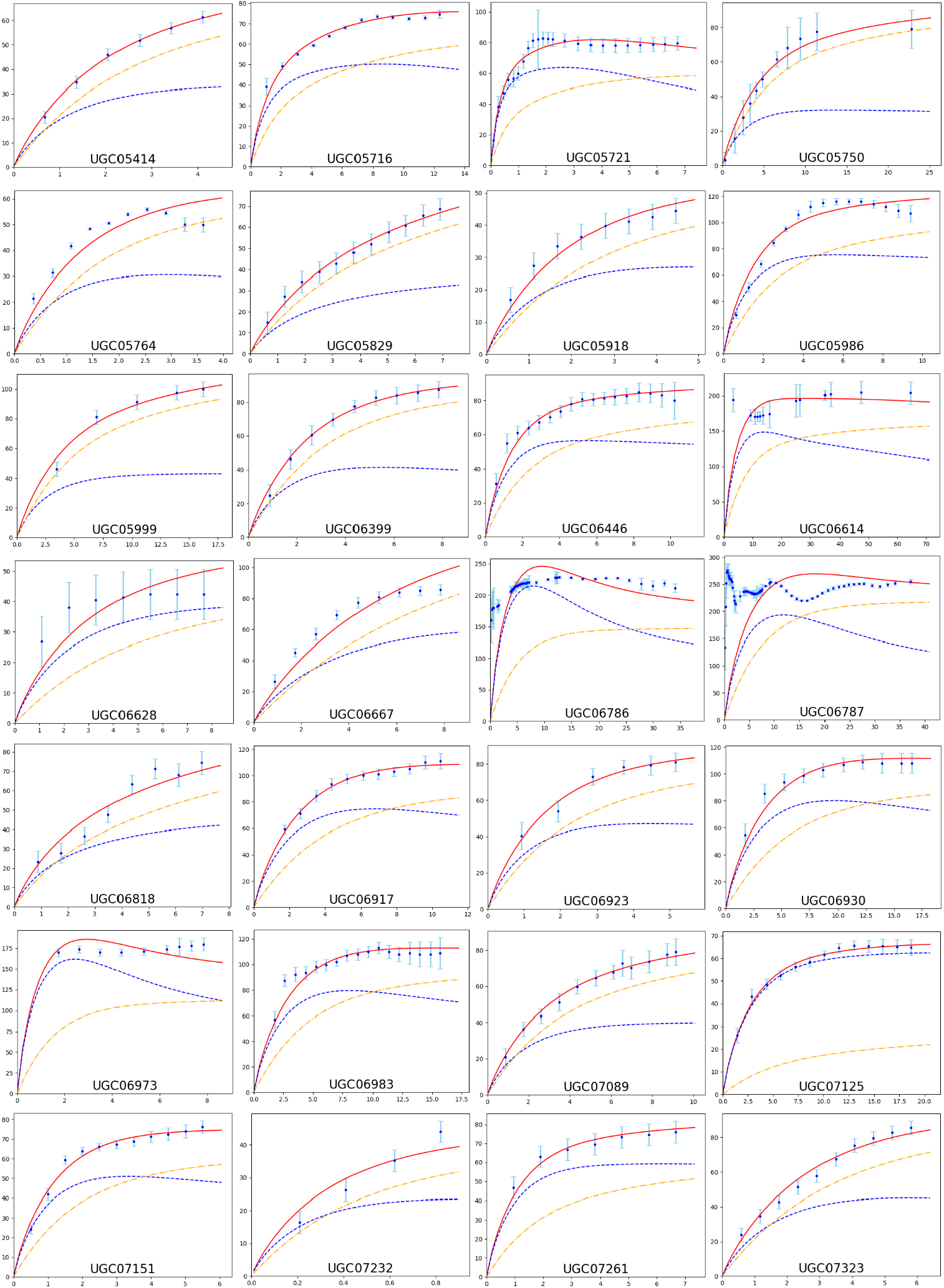}
\end{figure}

\begin{figure}[H]
	\centering
	\includegraphics[width=\linewidth]{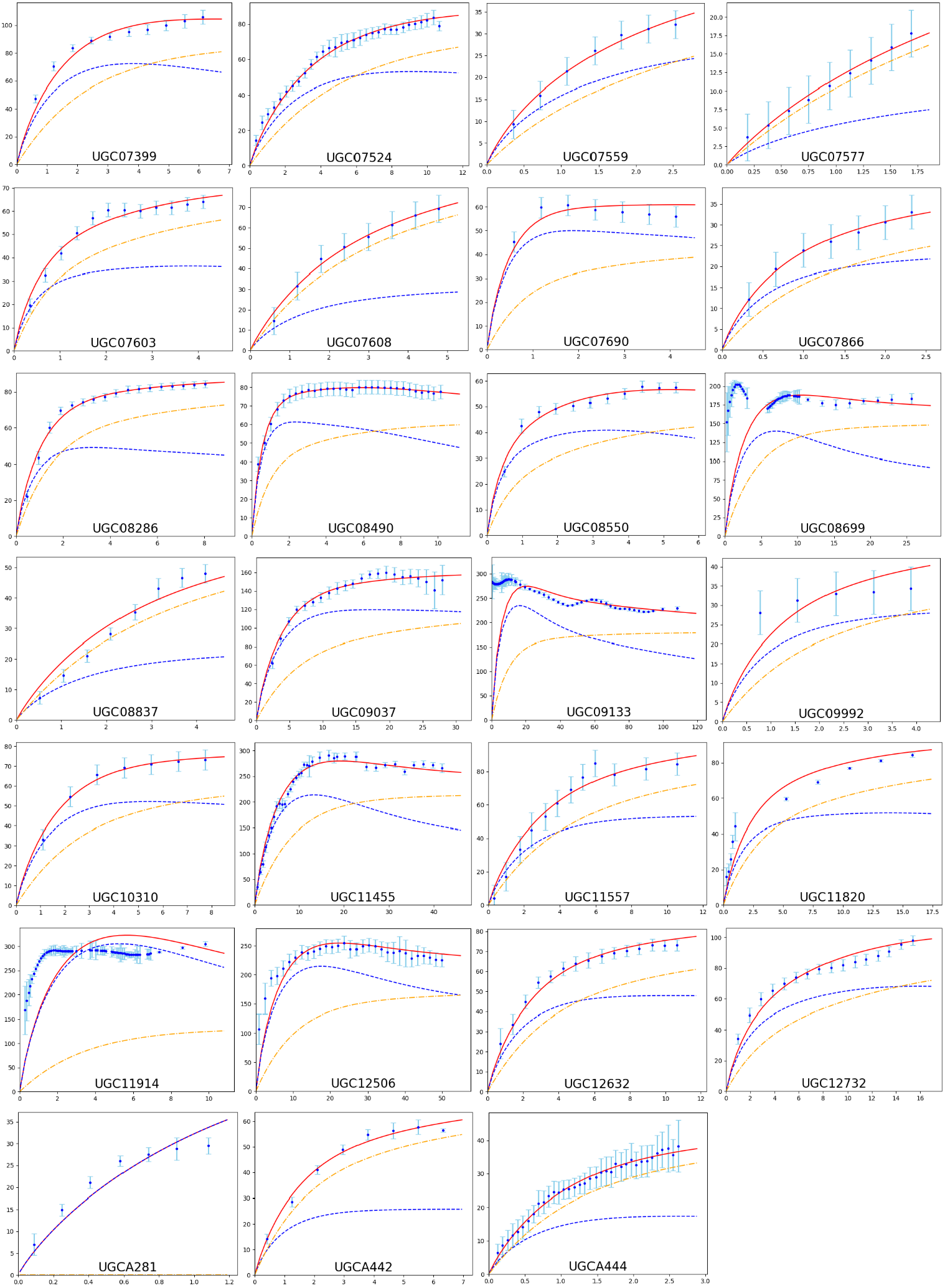}
\end{figure}

\end{document}